# Cyber Threat Hunting: Non-Parametric Mining of Attack Patterns from Cyber Threat Intelligence for Precise Threats Attribution

**Rimsha Kanwal[1], Umara Noor[2], Zafar Iqbal[3], Zahid Rashid[4]**

[1]Department of Computer Science, Faculty of Computing and Information Technology, International Islamic University, Islamabad, Pakistan
[2]Department of Software Engineering, Faculty of Computing and Information Technology, International Islamic University, Islamabad, Pakistan
[3]Department of Computer Science, National University of Computer and Emerging Sciences (NUCES), Islamabad, Pakistan
[4]Technology Management Economics and Policy Program, College of Engineering, Seoul National University, 1 Gwanak-Ro, Gwanak-Gu, 08826 Seoul, South Korea
rimsha.mscs1079@iiu.edu.pk, umara.zahid@iiu.edu.pk, zafar.iqbal@isb.nu.edu.pk, rashidzahid@snu.ac.kr

## Abstract

With the ever-changing landscape of cyber threats, identifying their origin has become paramount, surpassing the simple task of attack classification. Cyber threat attribution gives security analysts the insights they need to device effective threat mitigation strategies. Such strategies empower enterprises to proactively detect and defend against future cyber-attacks. However, existing approaches exhibit limitations in accurately identifying threat actors, leading to low precision and a significant occurrence of false positives. Machine learning offers the potential to automate certain aspects of cyber threat attribution. The distributed nature of information regarding cyber threat actors and their intricate attack methodologies has hindered substantial progress in this domain. Cybersecurity analysts deal with an ever-expanding collection of cyber threat intelligence documents. While these documents hold valuable insights, their sheer volume challenges efficient organization and retrieval of pertinent information. To assist the cybersecurity analyst activities, we propose a machine learning based approach featuring visually interactive analytics tool named the Cyber-Attack Pattern Explorer (CAPE), designed to facilitate efficient information discovery by employing interactive visualization and mining techniques. In the proposed system, a non-parametric mining technique is proposed to create a dataset for identifying the attack patterns within cyber threat intelligence documents. These attack patterns align semantically with commonly employed themes ensuring ease of interpretation. The extracted dataset is used for training of proposed machine learning algorithms that enables the attribution of cyber threats with respective to the actors. Experiments demonstrate that our approach attained cyber threat attribution with 95.35% accuracy, 95.56% recall, 96.20% precision, 0.93 F1-score, and a 2% false positive rate. Furthermore, the experiments reveal that usability of CAPE achieves notable average precision rates of 88.25% for document retrieval, and 88.8% for retrieval of cyber threat actor's information. The proposed approach will enable security analysts to perform more accurate and faster threats attribution compare to manual analysis.

## 1. Introduction

Attack patterns shows the behavior of cyber threat actors, also known as Tactics, Techniques, and Procedures (TTPs). By profiling a specific threat actor, we can identify common attack patterns shared across their attack campaigns. These attack patterns can take different forms including the use of distinct fake identities, installation of hidden traps on victim machines, and specific network activity patterns. The rise of Advanced Persistent Threats (APTs) demonstrates that most large-scale attacks aren't random but involve orchestrated campaigns with consistent patterns. Therefore, it is important to understand how attackers operate in order to implement strong defenses against them. However, pinpointing the strategic underpinnings of an attack proves challenging due to the sophisticated hacking techniques used by APT actors to target victims [1]. Cyber threat attribution is the process of identifying and assigning responsibility to threat actors (i.e., individuals, groups, or other entities) behind cyber-attacks. The process involves gathering and analyzing evidence from various sources, such as digital forensics, cyber threat intelligence (CTI), and behavioral analysis, to determine the origin, motives, tactics, techniques, and procedures (TTPs) of the attackers.

Recognizing cyber criminals based on their behavioral traits is very effective for informed decision-making. As the frequency of cyber-attacks rises, there is a simultaneous growth in the volume of data related to these incidents. A significant portion of this data is found in unstructured textual documents, making it challenging for cybersecurity analysts to interpret effectively. Organizations are increasingly using unstructured evidences to extract valuable insights and apply them in defending against cyber-attacks. This information is valuable for subsequent investigations by entities or legal bodies. The sharing of cyber threats information is a promising approach to stay ahead of security threats and prevent them proactively [2]–[4]. Having up-to-date information related to ongoing attacks and potential threats is crucial for effective preparation, providing an overall situational awareness of cybersecurity [2]–[5].

The rise in cybersecurity incidents leads to an exponential increase in CTI documents. Manual collecting and reviewing diverse, unstructured CTI documents is becoming increasingly challenging. The identification of attack patterns is helpful in minimizing the frequency of cyber-attacks. To attain a comprehensive understanding of the cyber-attacks landscape, cybersecurity analysts need to investigate past attack patterns of threat actors within their relevant scope. Several approaches for attributing cyber threats are available in the literature, utilizing various searching, clustering, and classification techniques (Section 2). However, these approaches still lack in accuracy, effectiveness and comprehensiveness.

To improve the accuracy of cyber threat attribution, it is essential to identify the clusters of keywords recurrently found within the CTI repositories, which are closely associated with attack patterns familiar to cybersecurity analysts (Section 3.3). Several clustering and topic modeling techniques are employed in the area of attributing cyber threats (Table 1). Some of these techniques are parametric and others are non-parametric, and have their own benefits and limitations (Section 2). Many existing methods are parametric, relying on specific parameters, underscoring the need for more robust techniques independent of such constraints (Table 1). Unlike keyword-based searches, non-parametric techniques have the advantage of effectively handling extensive textual thematic topics in theme-based searches [6]. Non-parametric techniques operate without specifying cluster numbers, offering enhanced flexibility (Section 3.3.3). It has been observed that there can be a co-occurrence of attack patterns within CTI documents, making non-parametric techniques more robust in handling such complexities [5]. Within CTI documents, a crucial aspect involves a unique non-parametric key term clustering, facilitating the identification of cyber-attack patterns through the co-occurrence of these patterns within the documents.

Advanced analytical approaches are crucial to assist analysts in navigating and retrieving pertinent attack patterns employed by cyber threat actors. Specifically, cybersecurity analysts require seamless access to relevant CTI reports aligned with their chosen attack patterns and an understanding of the evolution of cyber-attack patterns over time. Additionally, comprehensive datasets are not readily available, and analysts face challenges in navigating and retrieving pertinent attack patterns used by cyber threat actors. Establishing relationships between attack patterns and cyber threat actors is also difficult. Analyst commonly rely on keywords for searching and identifying attack patterns. However, basic searches using repository-provided keywords often lack advanced search capabilities due to the abundance of unstructured CTI repositories containing these terms. Furthermore, there is a need for improved visualization of search results to display trends and relationship between attack patterns of threat actors and the evolution of cyber-attack patterns over time, thereby, facilitating easier interpretation and accurate cyber threat attribution.

In this study, we propose a comprehensive approach of cyber threat attribution that achieve a high level of accuracy. We created a dataset by collecting CTI documents from multiple information sources (Section 3.1), and utilized for the identification of cyber threat actors. Our approach leverages non-parametric techniques to identify cyber-attack patterns by recognizing significant keywords in CTI documents (Section 3.2). For identification of attack patterns, our approach includes a visually interactive analytics tool called the Cyber-Attack Pattern Explorer (CAPE) to facilitate efficient investigation and information discovery through the utilization of interactive visualization and mining techniques. The clustering of attack patterns is done utilizing the non-parametric medoid-seeking algorithm (Table 3). The identified attack patterns are used to train the machine learning (ML) for cyber threat attribution. CAPE assists security analysts in gathering information related to attackers and their attack patterns. One promising advantage of CAPE lies in its ability to generate a list of cyber threat actors related to keywords entered by the analysts. The CAPE search bar enables interactive searches on themes, producing a list of prominent cyber threat actors utilizing these attack patterns. Additionally, CAPE facilitates the examination of the evolution of cyber threat actors over time through interactive visualization, exploration of relevant attack patterns, and the display of related cyber threat actors along with CTI documents.

This study makes a unique contribution by introducing a hybrid technique of cyber threat actor attribution that involves both clustering and classification. In the identification of cyber-attack patterns within CTI documents, we utilize a non-parametric medoid-seeking clustering technique, enabling the soft assignment of patterns to multiple documents. Our approach involves medoid seeking from density distribution, utilizing a density kernel to cumulatively estimate word densities. Additionally, a frequent pattern mining algorithm (FP growth) facilitates clustering by ranking essential words based on their frequency, using an adapted Term Frequency-Inverse Document Frequency (TF-IDF). The cyber threat actor's tactical attack pattern dataset, obtained using the proposed non-parametric approach, is then utilized to train five machine learning models developed for attributing cyber threat actors (Section 3.4). These machine learning models include (1) Random Forest - RF, (2) Decision Tree - DT, (3) K-Nearest Neighbors - KNN, (4) Deep Neural Networks - DNN, and (5) Naïve Bayes - NB. The results reveal that the combination of Naïve Bayes with the non-parametric approach attained 95.35% accuracy, 95.56% recall, 96.20% precision, 0.93 F1-score, and 2% false positive rate. Furthermore, the usability of CAPE is assessed by conducting usability experiments with 20 participants. The results show that the usability of CAPE achieves notable average precision rates of 88.25% for document retrieval and 88.8% for the retrieval of cyber threat actor information.

The presented approach of cyber threat attribution has multifaceted implications. The practical implication includes the provision of a comprehensive approach for cybersecurity analysts in the identification and exploration of tactical attack patterns using the CAPE tool. By integrating CAPE with a Security Information and Event Management (SIEM) system, the cybersecurity analysts can conduct in-depth analyses of cyber-attack tactics and attribute them to potential threat actors with high precision. Furthermore, the theoretic implication of our proposed approach includes devising a hybrid technique that integrates clustering and classification, thereby enriching the theoretical foundations of cyber threat actor attribution.

The rest of the paper is structured as follows: Section 2 provides an overview of state-of-the-art cyber threat attribution approaches and compares them with our proposed approach. Section 3 details our proposed approach of cyber threat attribution. Section 4 discusses the usability and performance analysis of the proposed approach. Section 5 presents the real-world case descriptions and implications of this study. Lastly, Section 6 concludes the paper, summarizing the findings and suggesting future directions.

## 2. State-of-the-art approaches for cyber threat attribution

Given the rising frequency of cyber-attacks, pinpointing the root causes is crucial for formulating effective countermeasures. Consequently, researchers are increasingly using machine learning algorithms for cyber threat attribution, aiming to develop automated defenses against these threats. This section explores the capabilities of state-of-the-art cyber threat attribution approaches that utilize various techniques, as outlined in Table 1.

To address the increasing volume of CTI documents, the adoption of topic modeling techniques is essential. These techniques, predominantly statistical, are designed to extract text patterns and themes from document collections. Once identified, these themes offer promising opportunities for statistical analyses of textual data [7]. One of the earliest examples of topic modeling is Latent Semantic Indexing (LSI), which applies matrix factorization to the word-document matrix to reveal themes and variations within textual data [8]. However, LSI categorizes documents using orthogonal topics, posing challenges in interpreting both positive and negative weight values assigned to keywords within topics [6]. In contrast, probabilistic topic modeling methods rely on nonnegative probability distribution functions across keywords and topics, resulting in more interpretable outcomes [7]. Notably, probabilistic Latent Semantic Indexing (pLSI) [9] and Latent Dirichlet Allocation (LDA) [10] have emerged as popular topic approaches, generating coherent topics by leveraging co-occurring words. However, pLSI faces issues such as overfitting and inefficiency, while LDA's topics may lack sufficient distinctiveness for effective document classification due to their proximity [11]. These techniques are characterized by their high computational demands [6]. Hierarchical Dirichlet (HD) methods [12], [13] have been introduced to extract dynamic topics from textual sources but require predefined hierarchy depths, often leading to the inclusion of numerous irrelevant stop words for cyber security analysts [6]. Nonnegative Matrix Factorization (NMF) [14], another matrix factorization-based technique, offers a nonnegative approach to topic modeling [15]. However, the consistency of NMF's results diminishes with an increasing number of topics [6]. Additionally, the evolutionary hierarchical clustering method generates hierarchical groups but may not be ideal for extracting attack patterns due to their shared characteristics [16]. As shown in Table 1, NMF based topic modeling has been used by [17] to analyze thousands of cybersecurity articles related to news and policymaking concerning national identities.

The research community has shown significant interest in word embedding, a technique that provides enhanced vector representations enriched with semantic nuances and word associations. Word embedding is a type of word representation technique used in Natural Language Processing (NLP) and Machine Learning (ML), where words are converted into numerical vectors within a continuous vector space. The key idea is to represent words in a way that captures their semantic relationships and contextual meanings. Essentially, words with similar meanings are mapped to vectors that are closely positioned to each other in the vector space. Notable examples of embedding models include Skipgram [18], Continuous Bag of Words [19], and GloVe [20]. These models typically use a limited window to gather contextual words situated near each target word. Recent studies have integrated NLP and Convolutional Neural Networks (CNN) to develop data-driven approaches for modeling and classifying tweets as related or not to cybersecurity. The cybersecurity-related tweets are further classified into a fixed list of cyber threats [21] (Table 1). Another study proposed an approach using NLP and ML to automatically identify and profile emerging cyber threats based on Open Source Intelligence (OSINT), aiming to generate timely alerts [22] (Table 1). Similarly, [23] employed NLP and ML algorithms to extract features from unstructured CTI reports and then attributing cyber threat actors (Table 1).

The text analysis techniques are traditionally complemented by visual analytic methods. For instance, TIARA [24] visualizes the flow of topical information in streaming text, while FacetAtlas [25] explores topic interconnections through graphical representation. TopicPanorama [26] depicts topic connections within document corpora, and ParallelTopics [27] visualized topic evolution over time using a thematic river and incorporates parallel dimensions for displaying document distribution across topics. TextFlow [28] facilitates graphical analysis of topic merging and splitting by tracking their evolutionary history. Additionally, Sun et al. [29] explored topical relationships through cooperative arrangements, while Literature Explorer [6] adeptly identifies topics without numeric values, focusing primarily on retrieving significant scientific publications based on specific topic queries rather than adapting the topics themselves.

Over the last decade, cyber threat attribution techniques faced significant efficacy limitations. These methods primarily relied on low-level Indicators of Compromise (IOCs), such as, IP addresses, file hashes, and domain names [30]–[32], without incorporating cyber threat actor's attack patterns [33], [34]. Some approaches were either manual or narrowly focused on identifying specific cyber-attacks [33], [34], while others restricted their analysis to a limited number of cyber threat actors [35]. Moreover, several models remained theoretical without empirical validation [36], [37]. Noor et al. [5] introduced a model aimed at automatically identifying attack patterns of cyber threat actors using Latent Semantic Indexing (LSI) (Table 1). They utilized data from ATT&CK MITRE [38] and CTI documents spanning events from May 2012 to February 2018, involving 36 cyber threat actors. Despite achieving a 94% accuracy rate, they acknowledged the limitations of LSI and the potential for false positives, without empirically comparing them to low-level IOCs for threat actor identification. In another study, Haddadpajooh et al. [1] proposed a model using fuzzy pattern trees, multi-modal fuzzy, and fuzzy c-mean partitioning, focusing on malware attribution and achieving 95.2% accuracy (Table 1).

Arun Warikoo [39] conceptualized a triangle model for cyber threat attribution, focusing on sectors, tools, and tactics, techniques, & procedures (TTPs) aiming to assist cybersecurity analysts. However, this model remained conceptual and lacked empirical validation. In another study, Naveen et al. [40] introduced a model by using neural network to perform cyber threat attribution, utilizing various CTI documents. This approach achieved an accuracy of 86.5%, but the dataset was limited to only 12 advanced persistent threats (APTs), and did not explicitly extract attack patterns from the CTI documents. They utilized the dataset which was originally published by Perry et al. [41]. Similarly, Sentuna et al. [42] introduced a model utilizing Naïve Bayes posterior probability to enhance both the prediction accuracy and processing speed of cyber threat attribution. The model achieved an accuracy of 95.28% but it was tested on a limited dataset of only 10 cyber threat actors. Additionally, Dionísio et al. [43] proposed a tool using CNN and Bidirectional Long Short-Term Memory (BiLSTM) Neural Network to collect, classify and extract information from security-related tweets (Table 1).

**Table 1** - Comparison of cyber threat attribution approaches

| **Ref.** | **Approach** | **Contribution in the area of cyber threats attribution** | **Domain** | **Reported Benefits** |
|---|---|---|---|---|
| "Noor et al., 2019 [5]" | Latent semantic analysis (LSA), Singular value decomposition (SVD) and ML algorithms | Proposed a framework to automate cyber threat attribution based on their attack patterns extracted from CTI reports. | Cyber threat attribution | This study proposes an automated mechanism to extract high-level indicators of compromise (IoCs) from unstructured CTI documents using a common vocabulary. The objective is to profile cyber threat actors and utilize these profiles for cyber threat attribution. The approach involves mapping attack pattern queries to their conceptual meanings rather than relying on keyword matches. The labels for high-level IoCs data set is prepared from the MITRE's Adversarial Tactics, Techniques, and Common Knowledge (ATT&CK) framework. |
| "Ghasiya and Okamura, 2021 [17]" | Nonnegative matrix factorization (NMF) modeling | Analyzed thousands of cybersecurity articles related to news and policy making of national identities. | Topic modeling | Using topic modeling techniques, this study examined the cybersecurity phenomena in six countries individually. The study includes a comparative analysis to identify potential correlations and shared patterns among these countries. |
| "Haddadpajouh et al., 2020 [1]" | Multi-view fuzzy pattern tree and multimodal consensus clustering technique | Propose a multi-view fuzzy consensus clustering model for attributing cyber threat payloads (malware) to its actor based on their different associated artifacts. | Attributing advance persistent threats (APT) malware groups | A customized sandbox was developed to extract views of APT malware. Using these views, a multi-view fuzzy consensus clustering model was built to attribute APT malware to their associated campaigns. Additionally, the study provided a comparison between single-view attribution and multi-view malware attribution. |
| "Irshad and Siddiqui, 2023 [23]" | Natural language processing (NLP) and machine learning (ML) algorithms | Extracting features from unstructured cyber threats intelligence (CTI) reports by using NLP and then attributing cyber threat actors by using machine learning algorithms. | Attribute or profile cyber threat actors (CTA) | A cyber-threat attribution framework is introduced for detailed and accurate profiling. This framework incorporates a novel embedding model, *attack2vec* trained on domain-specific datasets. |
| "Behzadan et al., 2018 [21]" | NLP and convolutional neural network (CNN) | Proposed a data-driven approach for modeling and classification of tweets as related or not to cybersecurity and classify the cyber-related tweets into a fixed listed of cyber threats. | Modeling and classification | This study curated an open-source dataset comprising of 21000 manually annotated cyber-related tweets. Additionally, they developed an open-source web application for annotation, exploration, and management of Twitter-based OSINT collections. This study employs a data-driven approach for detecting and classifying cyber-related tweets. |
| "Le et al., 2019 [44]" | Novelty detection model | Proposed a framework that learns form cyber threat intelligence (CTI) descriptions published in public repositories such as Common Vulnerabilities and Exposures (CVE) and classify unseen tweets as either normal or anomalous to CTI. | Classification | This study focuses on collecting and analyzing Twitter data. It proposes an automated framework for detecting cyber threat tweets using novelty classification. Additionally, the study includes an analysis to establish relationships between correctly classified tweets and threat descriptions in the CVE database. |
| "Dionísio et al., 2019 [43]" | Convolutional neural network (CNN) and bidirectional long short-term memory (BiLSTM) neural network | Proposed a tool to collect tweets from a selected subset of accounts, and discards unrelated tweets using keyword filtering. | Classification | This study proposes an end-to-end threat intelligence tool based on deep neural networks, eliminating the need for feature engineering. The tool processes cybersecurity information received from Twitter. Initially, a CNN identifies tweets containing security-related information. Subsequently, a BiLSTM model extracts named entities from these tweets to generate security alert or populate indicators of compromise. |
| "Marinho and Holanda, 2023 [22]" | NLP and ML | Proposed an approach to automatically identify and profile emerging cyber threats based on OSINT (Open Source Intelligence) in order to generate timely alerts. | Identification and profiling | This study proposed an approach to characterize or profile the identified threats based on their intentions or goals, offering additional context on threat landscape and suggesting potential avenues for mitigation. |
| This Study | Non-parametric medoid-seeking clustering and ML Algorithms | We propose a machine learning-based approach that utilizes a non-parametric method to identify cyber-attack patterns by recognizing significant keywords in CTI documents. | Cyber threat attribution and Clustering | This study proposes a practical approach for identifying, exploring and attributing attack patterns with cyber threat actors. It includes a novel searching tool that employs interactive visualization and mining techniques facilitating in-depth analyses of cyber-attack tactics and attribute them to potential threat actors with high precision. The proposed hybrid approach integrates clustering and classification for precise cyber threat actor attribution. |

A significant amount of information about cyber threat actors and their attack patterns are available in unstructured textual documents, posing challenge for cybersecurity analysts to interpret effectively. The distributed nature of this information complicates accurate and efficient cyber threat attribution. The MITRE ATT&CK framework serves a knowledge base for attributing cyber threat actors [45]. It is developed manually by cybersecurity experts at MITRE, a not-for-profit organization. This framework provides a structured approach to understanding and categorizing adversarial TTPs. This framework is widely used in industry by cybersecurity professionals.

While offering several benefits, the MITRE ATT&CK framework has some limitations. It provides a static catalog of TTPs that may not fully adapt to the dynamic nature of cyber threats or incorporate contextual adaptation. The framework primarily focuses on TTPs obtained after a cyber-attack. Its detailed and extensive nature can overwhelm smaller organizations with limited cybersecurity resources, leading to potential information overload due to the vast amount of information and techniques listed. The ATT&CK framework is a valuable resource, but implementing it in operational settings is resource-intensive and complex. Moreover, the knowledge base contains examples that are prone to bias. To mitigate these limitations, organizations should supplement the ATT&CK framework with additional resources and strategies tailored to address the evolving threats, initial access vectors, and non-technical vulnerabilities. This integrated approach is crucial for establishing a robust and adaptive cybersecurity posture.

The existing approaches discussed in (Table 1) are contributing to profiling and attributing cyber threats actors from various perspectives, utilizing various type of datasets and techniques. However, each approach has its limitations, such as the non-availability of comprehensive datasets, difficulties in navigating and retrieving pertinent attack patterns used by cyber threat actors, and challenges in establishing relationships between attack patterns and cyber threat actors. Other limitation includes automated searching capabilities, improved visualizations, seamless access to relevant cyber threat intelligence reports aligned with chosen attack patterns, adaptation to the evolution of cyber-attack patterns over time, and ease of interpretation with high accuracy for cyber threat attribution. In this research work, we propose a practical approach to automate the process of cyber threat attribution with high accuracy, aiming to overcome limitations of existing approaches.

## 3. Proposed approach of cyber threat attribution

To address the challenges of cyber threat attribution, there is a need to develop improved techniques that can effectively extract attack patterns from unstructured CTI documents. In this regard, we propose an approach for retrieving cyber threat actors' attack patterns from such CTI documents, utilizing a non-parametric thematic topic detection method. The extracted attack patterns are than utilized to train five machine learning models for cyber threats attribution. Subsequently, the performance of proposed models is evaluated. The framework comprises of following phases, as depicted in Figure 1.

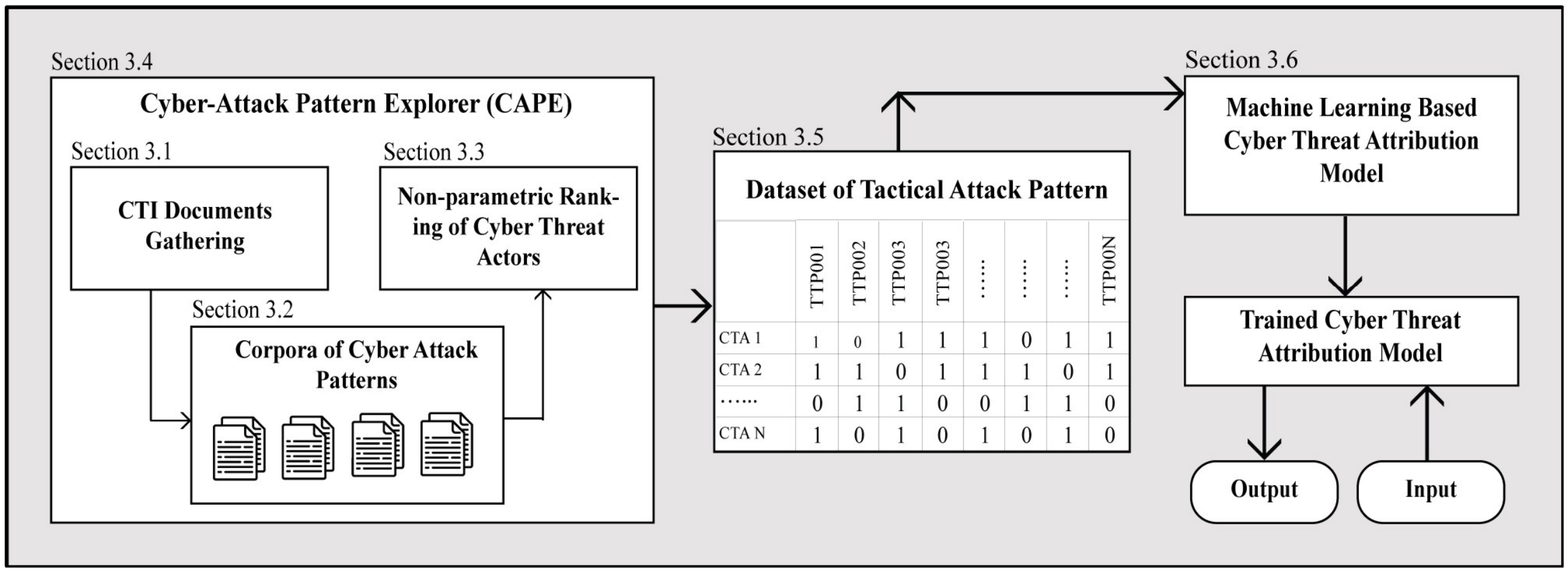

**Figure 1** - Proposed approach of cyber threats attribution

In the first phase, data is gathered in the form of unstructured CTI documents from various information sources, to build a corpus that captures patterns of cyber-attacks. The output of this phase is the corpus of CTI documents tailored to cyber-attack's patterns. In the next phase, data analytics is used to extract the attack patterns of cyber threat actors, resulting in a dataset organized as a "cyber threat actor – tactics, techniques & procedures (CTA-TTP)" matrix. Finally, the CTA-TTP matrix dataset is used to train machine learning algorithms, which are then applied to attribute cyber threats. The following sections provide detailed discussions of each phase of the proposed approach.

## 3.1 Data Collection

In this phase, data related to cyber threat actors were collected from trustworthy information sources in the form of unstructured textual CTI documents. These CTI documents, which describe the cyber threat incidents, also include attack patterns of cyber threat actors that are extracted in later phases. For this study, fifty cyber threat actors are identified from CTI documents. CTI documents are collected by using a customized search engine [46]. This search engine is specifically designed to find cyber threat actors, advanced persistent threats, and various malware. We also used another publicly available repository to collect CTI documents [47]. Using these information sources, we collected more than 726 CTI documents. The collected CTI documents contain information related to 50 cyber threat actors for conducting cyber threat attribution. The corpora include varying numbers of CTI documents relevant to each cyber threat actor. The primary difficulty we encountered during the collection of CTI documents was that different information sources used various names for the same cyber threat actor. The collected CTI reports cover incidents from May-2012 to May-2022, reported by various information sources in unstructured textual format.

These CTI documents encompass information related to different cyber threat actors and are extracted from different sources shown in Table 2. Notable sources of CTI documents include the U.S. Computer Emergency Readiness Team (US-CERT), security research websites, security vendors, security news websites, Indicators of compromise (IOC) miners, security blogs, and online forums comprising cybersecurity experts. The descriptions of these sources are provided in Table 2.

**Table 2** - Sources of cyber threat intelligence documents

| Sr. | Sources | Description | References (Examples) |
|---|---|---|---|

| No. | | | |
|---|---|---|---|
| 1. | Cybersecurity Expert Forum | Offer a valuable platform for professionals to connect, share knowledge, discuss challenges, and stay updated on the latest trends. | [48]–[51] |
| 2. | Security Blog | Online platforms where individuals or organizations share insights, news, and opinions on various security topics. These are valuable resource for staying informed about the latest threats, learning best practices, and connecting with other security professionals. | [52]–[55] |
| 3. | Security news website | Security news websites serve as essential resources for staying updated on the latest vulnerabilities, breaches, cybercrime trends, and industry developments. It enables individuals and organizations to understand the evolving threat landscape, make informed decisions about their security posture, and proactively prepare for potential attacks. | [56]–[58] |
| 4. | IOC miners | Tools or techniques used to automatically find and collect Indicators of Compromise (IOCs) from various sources. | [59] |
| 5. | Security research website | Provide valuable information, updates, and resources related to cybersecurity. | [49] |
| 6. | US-CERT portal | Web-based collaborative platform within the Cybersecurity & Infrastructure Security Agency (CISA) that facilitates information sharing between government and industry members regarding cyber threats and vulnerabilities. | [60] |
| 7. | Security vendors | Companies or organizations that offer products and services designed to help individuals, businesses, and governments protect their systems and data from cyberattacks and threats. | [61]–[63] |

## 3.2 Corpora of cyber-attack patterns

The collected corpora in Section 3.1 include more than 726 CTI documents. These documents are used to extract information such as attack patterns, cyber threat actors, victims, and campaigns. The CTI document contains high-level IOCs, a victim of the attack, the attacker, attack patterns, and malware used. Most of the CTI documents are in unstructured format and cannot be readily analyzed by the machines. Therefore, there is a need to process these CTI documents' corpora so that useful information can be extracted from them and convert into a format that can be easily processed and analyzed by machines. The CTI documents contain information related to attack patterns, which are described in various ways. The heterogeneous descriptions of attack patterns in CTI documents should have a clear and visible association, aiding security analysts in proper comprehending information and facilitating cyber threat attribution. Furthermore, the CTI documents also need to be clustered around the identified attack patterns. For clustering, we proposed a non-parametric approach to cluster the CTI documents of the corpora. We used the attack patters defined by MITRE ATT&CK [38] for clustering the CTI documents (Section 3.3.1).

The next step is to identify the attack patterns from CTI documents. The CTI documents in the corpora are collected from various sources and contain a variety of unstructured textual descriptions. Security experts from different organizations investigate cyber incidents and generate CTI reports. These documents explain the incident in terms of the purpose, motivation behind the attack, and capabilities of attackers, origin of attackers, and their connection with different states. Since CTI documents are generated by security experts at different organizations, they do not contain standardized keywords. To automatically detect these high-level IOCs from unstructured CTI documents, a nonparametric-thematic topic detection technique is proposed (Section 3.3.3). In this technique, rather than conducting a traditional keyword-based search, a new non-parametric thematic topic detection approach is proposed to detect the attack patterns from unstructured documents. To organize high-level IOCs, either a common descriptive language needs to be established, or an existing taxonomy may be utilized. MITRE ATT&CK [38] provides such a taxonomy, which can be readily used for thematic searching in corpora. This taxonomy is selected because it is based on various information sources and is continuously updated by the researchers at MITER ATT&CK.

## 3.3 Non-parametric ranking of cyber threat actors

In this section, a non-parametric approach is proposed to search for attack patterns within the unstructured textual CTI documents. The output of this phase is cyber threat actor profile matrix, which is utilized in the subsequent stages to train five machine learning models for effective attribution of cyber-attacks. This stage is divided into four major activities: keyword ranking, feature representation, non-parametric thematic topic detection and calculation of raking and relevance of cyber threat actors, as shown in Figure 2. This model is then applied to predict unseen data. A detailed discussion of each phase is provided in the following subsections.

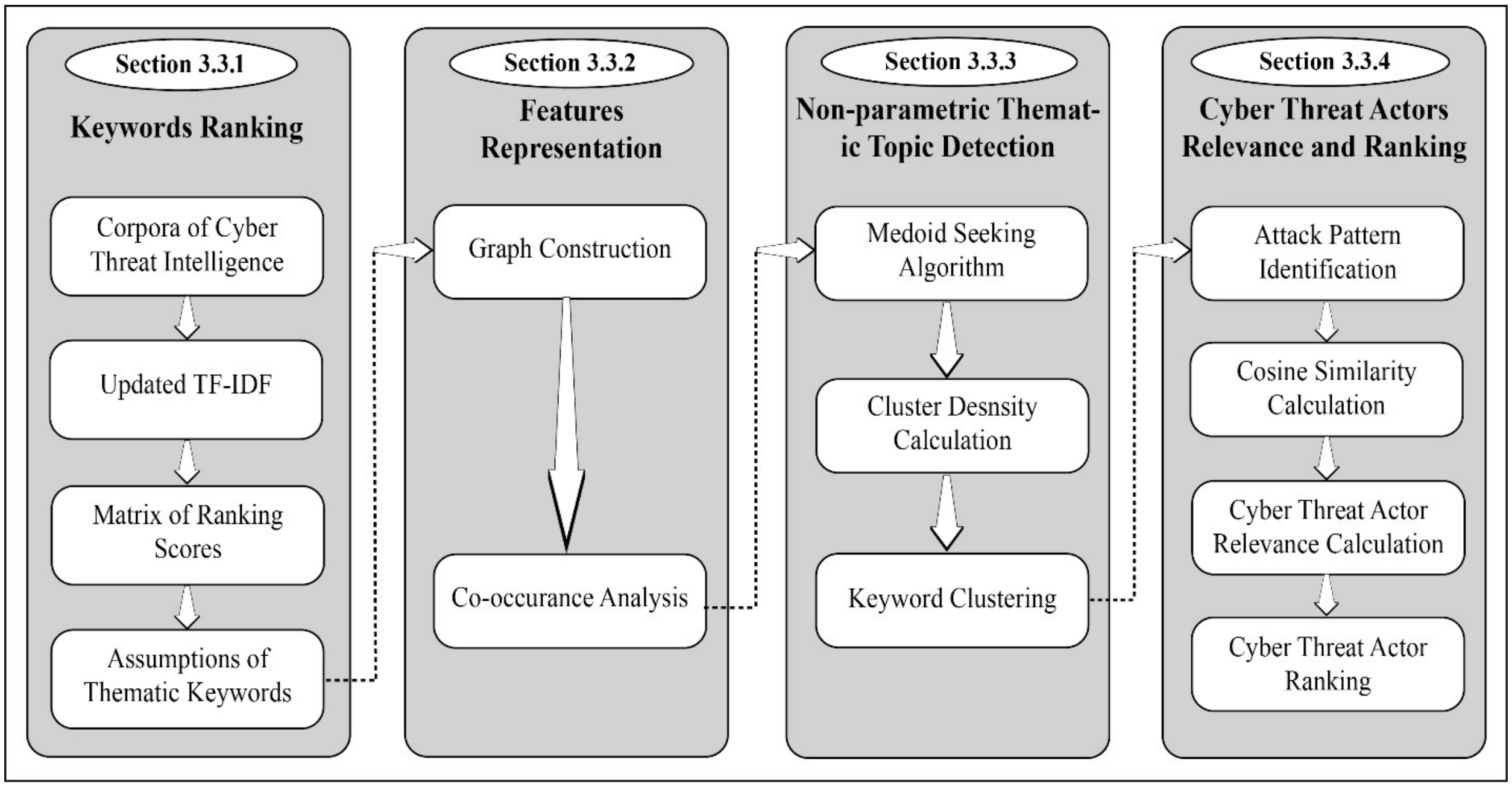

**Figure 2** - A non-parametric thematic topic detection-based approach for ranking cyber threat actors

### 3.3.1 Keyword ranking

The information related to cyber threat actors and attack patterns is gathered from various sources and linked to the cyber threat actors, as discussed in Section 3.1. These CTI documents contain crucial information such as attack patterns, cyber threat actors, victims, and cyber-attack campaigns. However, these CTI documents are textual and cannot be immediately analyzed by machines. The accumulated CTI documents provide insights into cyber threat incidents and associated actors. Collecting this data presents challenges due to the varying nomenclature used by different security companies for a single cyber threat actor. For instance, CrowdStrike [64] labels a particular cyber threat actor as "Comment Crew", Dell's

Secureworks [65] identifies them as "TG-8223", Mandiant FireEye [66] designates them as "APT1", and iSight refers to them as "BrownFox".

In the keyword ranking, an updated TF-IDF strategy is used to extract relevant words from a collection of CTI documents and evaluate their value based on overall frequency in the entire corpora. This method analyzes the fraction of documents in the entire corpora that include these words. This approach relies on two assumptions:

(a) Important thematic words appear frequently
(b) Important thematic words may not be present in all documents from the field of cybersecurity

Based on the assumptions, a ranking score matrix $\pi\ (wd, dc)$ is created. This matrix describes the ranking of every word '*wd'* in the textual CTI document '*dc*'. The total score $\pi\ (wd)$ of the word '*wd'* is as follows:

$$\pi\ (wd, dc) = (fr\ (dc, wd) \times idf\ (wd) \times ir\ (wd)\ ) \tag{1}$$

$$\pi\ (wd) = P_{\ |D|}\ \pi\ (wd, dc) \tag{2}$$

In equation 1, $fr\ (wd, dc)$ denotes the frequency with which a word *wd* occurs in a textual CTI document *dc*. In addition, the inverse document frequency of word *wd* is represented by $idf\ (wd) = n\ (wd)\ /\ Nd$, where $n\ (wd)$ represents the number of CTI documents that contain the word *wd*, and *Nd* represents all CTI documents in the corpus. Words with low frequency are removed by using $ir\ (wd)$. The value of its parameter *P* is set experimentally, as shown in equation 3.

$$i\ r(wd) = \begin{cases} 1\ if\ n(wd) \geq P \\ \frac{n(wd)}{P}\ Otherwise\ n(wd) < P \end{cases} \tag{3}$$

In this work, we used the term ranking in non-parametric thematic topic detection.

### 3.3.2 Feature representation

In the feature representation, the extracted keywords discussed in section 3.3.1 from the documents are used as variables. Subsequently, an undirected graph GR = (VN, EK) is used to model the relationship between these keywords. Here, VN represents the extracted keywords, and EK denotes the edges connecting these keywords. The graph illustrates word connections based on their co-occurrence in the document corpus; if two keywords co-occur within the corpus, they are linked in the graph. The strength of these connections is described as follows:

$$\sigma(wd_s, wd_t) = |\ \{dc_i\ |\ wd_s \in DC_i,\ wd_t \in dc_i,\ dc_i \in C_r\}\ | \tag{4}$$

The equation 4 describes the relationship between the terms $wd_s$ and $wd_t$. The number of all elements is denoted by using the symbol "| |". The document and corpus are represented by using $dc_i$ and $C_r$ respectively. This indicates that the words $wd_s$ and $wd_t$ exist in these documents. The FP-growth method is used to collect the co-occurrence of words. The divide-and-conquer approach in FP-growth efficiently extracts common patterns without requiring the generations of candidate sets.

### 3.3.3 Non-parametric thematic topic detection

The subsequent challenge involves detecting these attack patterns within CTI documents. These textual documents provide a diverse range of descriptions, lacking precise keywords aligned with the MITRE ATT&CK taxonomy. Instead of conventional keyword-based searches, we used non-parametric thematic topic detection approach to identify attack patterns within unstructured documents. This approach relies on thematic topics commonly used by security experts and analysts.

A non-parametric medoid-seeking algorithm is used to cluster the data, allowing soft clustering where a word may belong to multiple clusters. Being a non-parametric technique, it does not require predetermined number of clusters as input. The method identifies modes from density distribution by using a density kernel to estimate the density of each word in the CTI documents. Density accumulates across neighboring words to select medoids as cluster centers (also known as modes). Initial density calculations are centered around these medoids, followed by iterative updates using cluster density and term-to-cluster probability. In the context of word space, a medoid is a point that is closest to the center of a set of sample points, minimizing distances to all other data points. Each medoid is derived from a single term and represents a distinct meaning directly associated with keywords, providing a robust representation compared to calculating mean data points.

The density at the word $wd_j$ is defined as an aggregation of the combined effects from the densities of other words, as illustrated in equation 5.

$$f(wdj) = \sum_{i=0}^{n} m(wd_i) \times f_i(d(wd_i, wd_j)) \quad (5)$$

In equation (5), all keywords are represented by n. The value $m\ (wd_i)$ can be obtained using value of $\pi\ (wd_i)$ from equation (2), which represents the importance score of keywords.

$$m(wd_i) = \frac{\pi\ (wd_i)}{\sum_{j=1}^{n} \pi(wd_j)} \quad (6)$$

The density function of probabilities, which depends on the distance $d$ between two words $wd_i$ *and* $wd_j$ is shown in equation (6). Specifically, the calculation of the density function of $wd_i$ towards $wd_j$ depends on the distance between these words. A Gaussian kernel is applied for the density calculations, with the kernel size adaptively determined based on data scatter plots. The probability of $wd_i$ is used for confinement purposes. Consequently, a word of greater importance exerts a stronger influence on its neighbors. Initially, a batch of medoids is identified by locating local maxima after calculating the density at each $wd_j$, resulting in a set of medoids $H_k$ as shown below:

$$H_k = \{wd_k \mid f(wd_k) > f(wd_j),\ \forall j \epsilon \Gamma(wd_k)\},\ k = 1, \ldots, K \quad (7)$$

Using this non-parametric method, medoids can be identified as local maxima or nodes with the highest density among immediate neighbors. The number of clusters depends on the count of these local maxima.

The keyword clustering is computed through iterations following the identification of medoids, which serve as cluster centers. This process involves determining the probability that a word $wd_i$ belongs to each cluster $H_k$, thereby constituting soft clustering. The initialization of this probability is based solely on the distance between $wd_i$ and $H_k$, as described below:

$$m\ (H_k \mid wd_i) = \frac{1 - d(H_k \mid wd_i)}{\sum_{k=1}^{K} 1 - d(H_k \mid wd_i)} \quad (8)$$

where $d(H_k, wd_i) = d(wd_{Hk}, wd_i)$.

The equation 8 represents the medoid of cluster $H_k$ where K denotes the total number of clusters determined by the local maximum. Initially, the probability of a word is assigned to various medoids based solely on standardized distance, and this assignment is iteratively updated. The initial probability calculation by using equation (10) along with the density cluster for every cluster $k$ at word $wd_j$. The total density contribution from each additional word at $wd_j$ is as follow:

$$T_k(wd_j) = \sum_{i=0}^{n} m(H_k \mid \mathrm{wd}_i) \times f_i(d(\mathrm{wd}_i, \mathrm{wd}_j) \times rank(H_k) \quad (9)$$

Using the FP-growth approach, frequent pattern mining is conducted to identify cyber threat actors by determining the co-occurrence of cyber-attack patterns within documents. To accommodate queries from more CTI documents, additional documents can be added to the system. This process needs to be performed offline only once, assuming a sufficient document in the corpus. When new documents are added to the system, they can be queried by cyber security analyst.

The employed non-parametric approach is used to cluster documents related to cyber threat actors. The attack patterns defined by MITRE ATT&CK are utilized to identify their presence within the clustered documents of cyber threat actors [45]. Because security experts describe attack patterns in diverse ways, this approach results in a collection of heterogeneous attack pattern descriptions. These descriptions, varying in presentation, are linked to the attack patterns and contribute to enhancing the understanding of information for security analysts and cyber threat attribution.

For the organization of attack patterns, either establish a common descriptive language or leverage an existing taxonomy, such as MITRE ATT&CK [45], which is particularly suitable for this purpose. MITRE ATT&CK's selection is based on its compilation from various cyber threat attack reports, where security experts investigate the attack's purpose, motivation, and the attackers' capabilities while linking them to state actors. This taxonomy is updated regularly by the MITRE and have information about 570 TTPs and approx. 700 malware / attack tools at the time of this study.

### 3.3.4 Cyber threat actor's relevance and ranking

The ranking of the documents for the specified attack patterns is based on this relevancy score. Consequently, a list of ranked textual CTI documents can be found based on their relevance to the attack patterns. In equation (9), $n$ represents the total number of words, and the initial importance ranking of $H_k$, which is set to 1 for all clusters, is represented by the expression rank ($H_k$). Using equation (11) shown below, this ranking is updated repeatedly. As a result, each word contributes a different amount of density based on how far it is from cluster $k$ to word $wd_j$. This also depends on how strongly these words correspond to the cluster $H_k$ and the probability that representing that fact. Standardized densities of each of $n$ words at cluster $k$ are used to compute $m\ (wd_i \mid H_k)$. In more detail, this process is carried out as follows:

$$m(wd_i \mid H_k) = \frac{\mathrm{T_k}(wd_i)}{\sum_{i=0}^{n} \mathrm{T_k}(wd_i)} \quad (10)$$

The importance rating of each cluster k is modified based on the significance of the words and the degree to which they are members of the cluster. This is shown in equation (11):

$$Rank\ (H_k) = \sum_{i=1}^{n} m(H_k \mid \mathrm{wd}_i) \times \pi(\mathrm{wd}_i) \quad (11)$$

Additionally, $m\ (H_k \mid wd_i)$ is updated by using standardized densities of all clusters which are shown as follows:

$$m(H_k \mid \mathrm{wd}_i) = \frac{\mathrm{T_k}(wd_i)}{\sum_{i=0}^{n} \mathrm{T_k}(wd_i)} \quad (12)$$

The procedures involving equation (9) to (12) are continued until the findings remain unchanged. This method identifies the words $wd_i$ that belong to each cluster $H_k$. The conditional probability $m\ (wdi \mid H_k)$ is used to rank the words $wd_i$, which collectively represent the attack pattern. Furthermore, the similarity between the attack patterns is calculated using their connected keywords. The conditional probability $m(wd_i \mid H_k)$ is used to weight each dimension that represent the attack patterns. Cosine similarity is used to calculate how similar they are. After the identification of attack patterns, the connection of a document to an attack pattern can be determined by evaluating how much the words are related to attack patterns:

$$m(H_k \mid dc) = \frac{\sum_{i=1}^{n} \mathrm{n}(wd_i, \mathrm{dc}) \times \mathrm{m}(H_k \mid wd_i) \times \pi(wd_i)}{\sum_{k=1}^{K} \sum_{i=0}^{n} \mathrm{n}(wd_i, \mathrm{dc}) \times \mathrm{m}(H_k \mid wd_i) \times \pi(wd_i)} \tag{13}$$

The above equation calculates the relevance between a document and attack patterns. In equation (13), $dc$ represents the documents, and $n(wd_i, dc)$ depicts the co-occurrence of words in the documents. Additionally, based on the attack pattern's relevance with words and the proportion of words in the attack patterns, the important documents are determined.

$$m(dc \mid H_k) \sim \frac{\sum_{i=1}^{n} \mathrm{n}(wd_i, \mathrm{dc})}{\mathrm{n}(\mathrm{dc}) \cdot \mathrm{m}(wd_i \mid H_k)} \tag{14}$$

The algorithm for the ranking of textual CTI is given in **Table 3.**

**Table 3** - Algorithm for ranking of words ($wd_i$)

| | |
|---|---|
| **Input** | A document corpus from a cyber threat intelligence repository |
| **Output** | A collection of words $wd_i$ ranked according to the conditional probability $m(wd_i \mid H_k)$ for each cluster $H_k$ |

```
Function nptTopicDetection (data)
    // Compute the accumulated density for each word wd_j using Equation (5)
    for each word in document do
        accDensity ← accumDensity (word)
    End
    for each k in H_k do
        initRank (H_k, 1)
    End
    initCondProbability ← calculateProbability (H_k, wd_i);
    for each cluster(k) do
        // Compute the cluster density for each cluster k at wd_j using Equation (9)
        clusterDensity ← calculateClusterDensity (k, wd_j );
        // Compute the conditional probability m (wd_i | H_k) using Equation (10)
        condProbability1 ← calculateProbability (wd_i , H_k);
        // Compute the ranking of each cluster k using Equation (11)
        clusterRank ← calculateRanking (k);
        // Update the conditional probability m (H_k| wd_i) using Equation (12)
        condProbability2 ← updateProbability (H_k, wd_i);
        // Repeat the loop until there is no change in p (wd_i | H_k)
        if no change in condProbability1 then
            break
        Else
            Continue
        End
    End
    return accDensity
End
```

### 3.4 Cyber-attack pattern searching using CAPE

The first phase involves sourcing authentic data related to cyber threat actors. Information about cyber threat actors is often dispersed across multiple sources and typically exists in unstructured textual format within CTI documents. These documents contain extractable attack patterns employed by threat actors, as well as crucial details such as attack patterns, victims, malware, software tools used, and campaigns. However, due to their unstructured nature, CTI documents cannot be readily analyzed by machines.

The distributed nature of CTI information requires a systematic approach to gather CTI documents from various sources and link them to cyber threat actors. To address this challenge, we proposed an interactive search system named CAPE, designed to collect and analyze CTI reports from multiple distributed sources. The main objective of CAPE is to assist cybersecurity analysts by automatically and efficiently collecting relevant information about cyber threat actors and their attack patterns. This system comprises of three main modules: the data repository, CTI analytics, and interactive visualization module as shown in Figure 3. The data repository collects unstructured data from various sources, which is then used for data analysis.

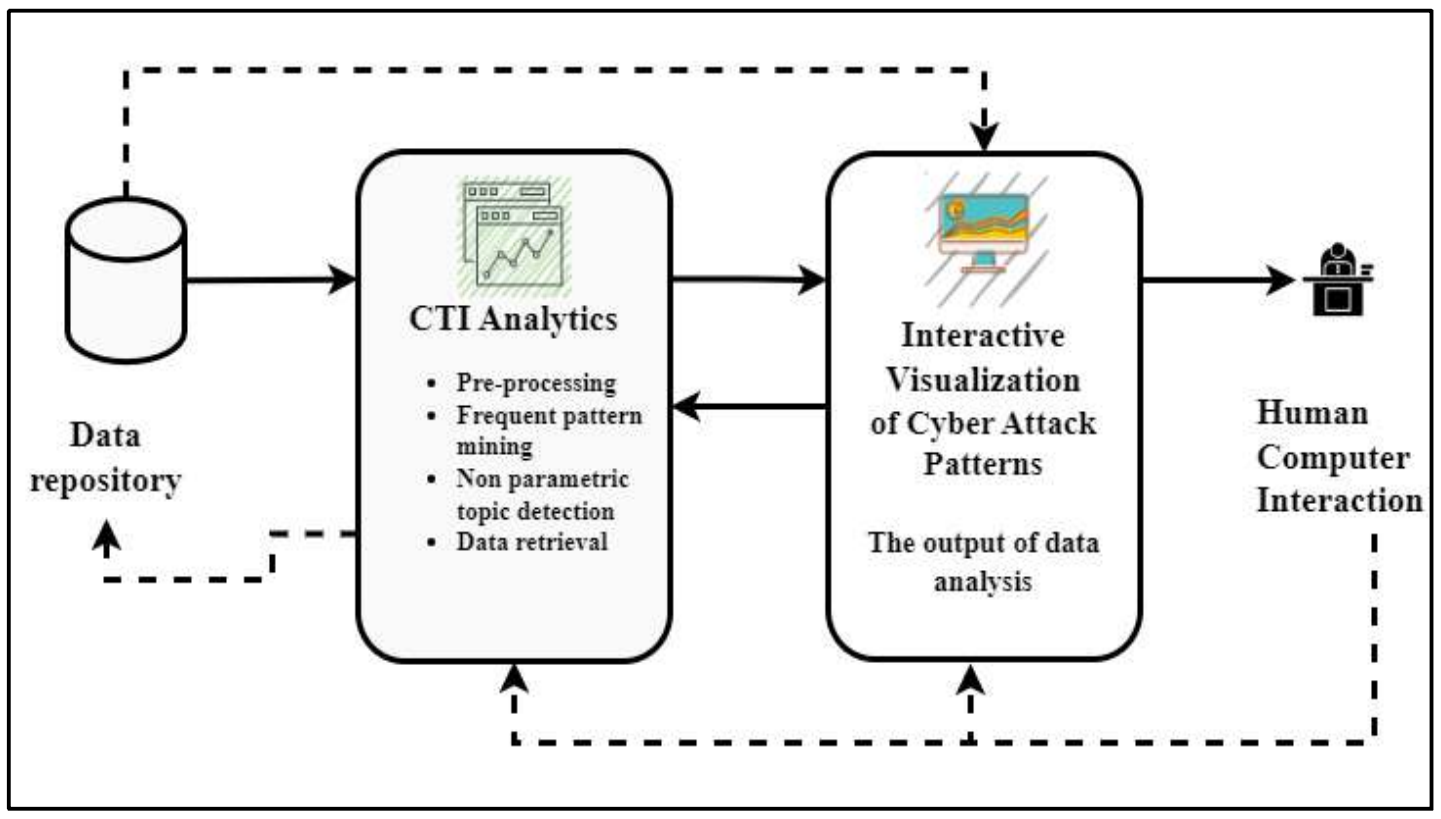

**Figure 3** - Overview of cyber-attack pattern explorer (CAPE)

The database stores data from each intermediate stage, with the final results accessible to cybersecurity analysts. Various stages are involved in pre-processing, including format conversion, cleansing (removing erroneous words), and information retrieval (such as cyber-attack patterns and cyber threat actors). Cleaned data is stored in the database for future decision-making process, while references are processed for future use. Additionally, word frequencies are calculated, and paragraphs and sentences are segmented based on these frequencies.

The data repository stores both the outcomes of data analysis and metadata about the CTI reports. Data analysis includes preprocessing of unstructured data, mining of frequent patterns, data retrieval, and non-parametric topic detection. The results of this data analysis are displayed in an interactive visualization.

In our proposed approach, during the initial stage of cyber threat attribution, cybersecurity analysts start by searching of CTI documents using attack patterns to identify cyber threat actors responsible for the attacks. The operational flow of CAPE search system is illustrated in Figure 4. The cybersecurity analyst requires interactive visualization to search attack patterns, identify cyber threat actors using these attack pattern, and obtaining brief description of these cyber threat actors. This system enables efficient investigation of cyber threat actors and their attack patterns. CAPE utilizes visual features directly related to data mining, providing straightforward information representation to enhance user comprehension and engagement. The system is scalable with the addition of more textual CTI documents. APIs facilitate communication and data sharing among the system's four components. For data queries, RESTful APIs and standard HTTP are used to establish connections between clients and servers. The data analysis module operates sequentially, utilizing a pipeline approach including preprocessing, frequent pattern mining, detection, and visualization of attack patterns

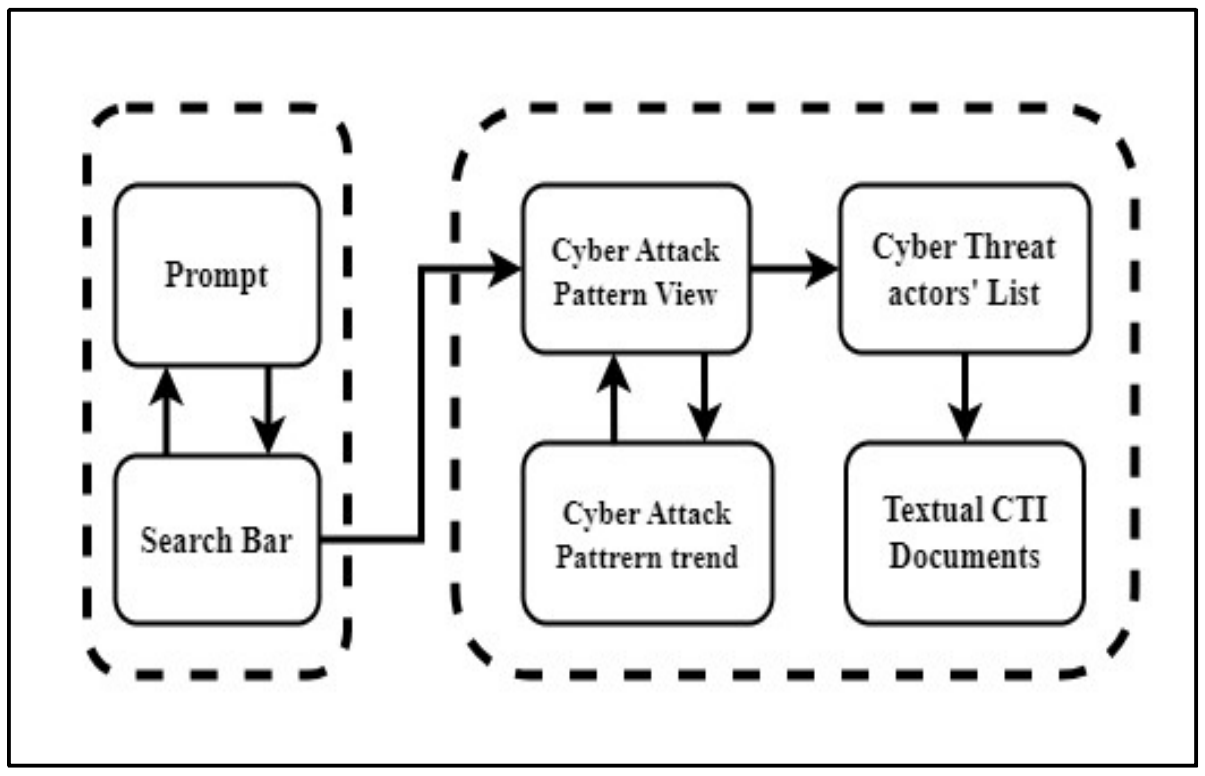

**Figure 4** - Operational flow of cyber-attack pattern explorer

## 3.5 Dataset of tactical attack patterns

The output of Section 3.4 is the CTA-TTP matrix (shown in Figure 5). In this matrix, each CTA (Cyber Threat Actor) and TTP (Tactics, Techniques, and Procedures) is represented with a unique ID attribute. The CTA is the target class to be predicted by machine learning model, while TTPs serve as the features or predictor attributes used as input for predicting the target class (CTA). The purpose of this matrix is to link CTAs with their relevant TTPs.

In the dataset, the presence or usage of a TTP by a CTA is represented by the value '1' or 'True,' while its absence is represented by '0' or 'False.' This CTA-TTP matrix dataset is then used to train the machine learning models (discussed in Section 3.6). The dataset includes 50 CTAs reported in CTI documents, resulting in 726 instances in the matrix dataset. The TTPs reported by ATT&CK MITRE are identified using CAPE in the CTI documents corpus. CAPE determines which TTPs are employed by a CTA, and these entries are recorded in the dataset used for training the machine learning model that attributes cyber-attacks to CTAs. These models are then used to predict the class of a CTA for an unseen cyber threat incident.

**Dataset of Tactical Attack Pattern**

| | TTP001 | TTP002 | TTP003 | TTP003 | …… | …… | …… | TTP00N |
|---|---|---|---|---|---|---|---|---|
| CTA 1 | 1 | 0 | 1 | 1 | 1 | 0 | 1 | 1 |
| CTA 2 | 1 | 1 | 0 | 1 | 1 | 1 | 0 | 1 |
| …… | 0 | 1 | 1 | 0 | 0 | 1 | 1 | 0 |
| CTA N | 1 | 0 | 1 | 0 | 1 | 0 | 1 | 0 |

**Figure 5** - Cyber threat actor - tactics, techniques & procedures (CTA-TTP) matrix dataset

## 3.6 Machine learning based cyber threat attribution model

The dataset is created in the section 3.5 is utilized to train five different machine learning models. These models predict the class of cyber threat actors for the unseen dataset. The five main machine learning algorithms used for prediction are Decision Trees (DT), Naïve Byes (NB), K Nearest Neighbors (KNN), Deep Learning Neural Network (DNN), and Random Forest (RF).

A decision tree is a type of machine learning algorithm used for both classification and regression tasks. DT is a tree-like structure where each internal node represents a "test" on a feature, each branch represents the outcome of the test, and each leaf node (terminal node) holds the resulting classification or prediction. The precision, readability, and stability obtain from DT makes it an important prediction model. Its ability to address problems with data-fitting, such as classification and regression, makes it useful for fitting non-linear connections.

A classification algorithm built on the Naïve Bayes' Theorem is known as a NB classifier. It is used for solving classification problems, with its main application as an efficient text classification. NB helps in the developing speedy machine learning models capable of making accurate predictions. The training phase requires only a small amount of data and can handle both discrete and continuous data. Its implementation is straightforward. NB is often well suited for real-time predictions compared to other machine learning algorithms because of its quick prediction time.

K-Nearest Neighbors (KNN) is one of the most fundamental machines learning algorithms, utilizing a supervised learning method. KNN examines the resemblance between the available data and new data, then classifies the new data into the group related to the available data. All data stored by the KNN algorithm is further used to classify new data based on similarity. Although KNN is most frequently employed for classification issues, it can also be utilized for regression problems. KNN is a non-parametric technique and is also known as a lazy learner because it does not instantly learn from the training set. Instead, KNN simply stores data when the training phase starts, then classifies new data into the group most similar to the stored data. KNN is simple to set up and comprehend, but it has the primary disadvantage of being substantially slower as the size of the dataset increases.

Random forest (RF) is a supervised machine learning algorithm designed to address classification and regression problems. RF uses ensemble learning, which integrates multiple classifiers to tackle complex issues. It is constructed using several decision trees trained via the bagging method, enhancing of accuracy of various machine learning models by aggregating predictions from multiple trees. RF assess model performance based on collective predictions made by these decision trees. Prediction accuracy of RF improves with an increase in the number of trees, and vice versa. RF addresses drawbacks of the DTs, such as overfitting, thereby enhancing model accuracy. It makes predictions without requiring explicit knowledge of features and provides an effective approach of handling missing data. The robustness of RF eliminates the need for hyperparameter tuning, allowing it to make predictions effectively with or without adjustments to hyperparameters.

Deep Neural Network (DNN) use of sophisticated algorithms and deep learning techniques. They simulate human brain activity by combining data inputs, biases, and weights. These components collaborate to effectively detect, categorize, and characterize entities in datasets. DNNs consist of multiple connected layers of nodes, with each layer enhancing prediction and classification. These interconnected nodes receive input, process it, and generate output. DNNs enable the handling of complex processes and can work with both structured and unstructured data which can be scaled as per requirements.

## 4. Usability and performance analysis of proposed approach

### 4.1 Usability analysis

The goal of our proposed approach is to assist cybersecurity analysts in collecting essential CTI reports through interactively selected keywords. To achieve this objective, we developed Cyber Attack Pattern Explorer (CAPE) based on our proposed approach. The user interface of CAPE is shown in Figure 6, facilitates the process of attribution of cyber threat actor. The visualization is implemented using "D3.js" [67], while the "Django" framework is utilized to make it a web-based application [68].

The main objective of this system is to assist the cybersecurity analysts in automatically and efficiently collecting the relevant information about cyber threat actors and their attack patterns. The system focuses on two main goals: simplicity and clarity. Firstly, it provides cybersecurity analysts with immediate access to information about cyber threat actors related and their cyber-attack patterns without manual effort. Secondly, it offers a comprehensive overview of cyber-attack patterns, cyber threat actors, a search bar, graphs, and the evolution of attack patterns over time, all presented with a clear design and user-friendly interface. CAPE is specially designed to enable cybersecurity analysts to search attack patterns effectively and efficiently, addressing a significant challenge posed by the rapidly increasing number of CTI documents. Consequently, there is a high demand for quick and effective access to attack patterns and threat actors. To meet this demand, a user-friendly interface is created with the following goals:

**Goal 1: Simplicity** in CTI document collection is achieved through interactive searches, allowing users to gather CTI documents related to specific cyber threat actors of interest by using specific keywords.

**Goal 2: Clarity** in understanding relevant attack patterns involves users comprehending the connections between the selected keyword, the list of cyber threat actors, and the associated documents.

**Goal 3: Clarity** in understanding the evolution of cyber-attack patterns enables users to track trends by observing the frequency of their use by various cyber threat actors.

The search bar serves as the main entry point for inserting attack pattern keywords. Here's a description of the primary usage scenario: cybersecurity analysts utilize the search bar to input cyberattack patterns. As user start typing, relevant attack patterns are suggested in a drop-down list. Users can then select the desired input from the provided list. Subsequently, the user interface dynamically updates to reflect the new selection, highlighting the targeted term with a red circle and displaying associated attack patterns connected in a graph view. Following that, the user can assess:

- The interactive graph view presents attack patterns related to the selected keyword.
- Cyber threat actors associated with selected attack patterns are displayed in the list on the right side. These actors can be sorted either in ascending order or descending order. In the description view, each actor can be selected to display a short description.
- The usage of attack patterns over time is represented using a graph.

The graph view employs a basic graph to illustrate the relationship between various keywords. Users can interactively select a keyword from the graph to further examine cyberattack patterns, cyber threat actors, and evolution of attack patterns over time.

The CAPE underwent evaluation by human participants using a collection of CTI documents obtained from a customized search engine. This database contained information specifically related to cyber threat actors and their tactics, encompassing a total of 726 complete paper documents. The evaluation aimed to showcase how the CAPE assists individual cybersecurity analysts in performing tasks within the cybersecurity domain. Ten participants of MS and Ph.D. levels actively engaged in evaluating the CAPE search process performance. They utilized the system to assess its effectiveness in areas aligned with their expertise, particularly in analyzing the CTI documents. The participants successfully completed the designated tasks outlined in usability surveys. Evaluation requirements, visual components, and methods are specified in Table 4. The evaluation results provide a rating for the approach with a focus on clarity and simplicity goals mentioned earlier. The precision rate, indicating the accuracy of system-retrieved documents compared to participants' knowledge, was calculated by determining the ratio of correct documents to the total retrieved documents.

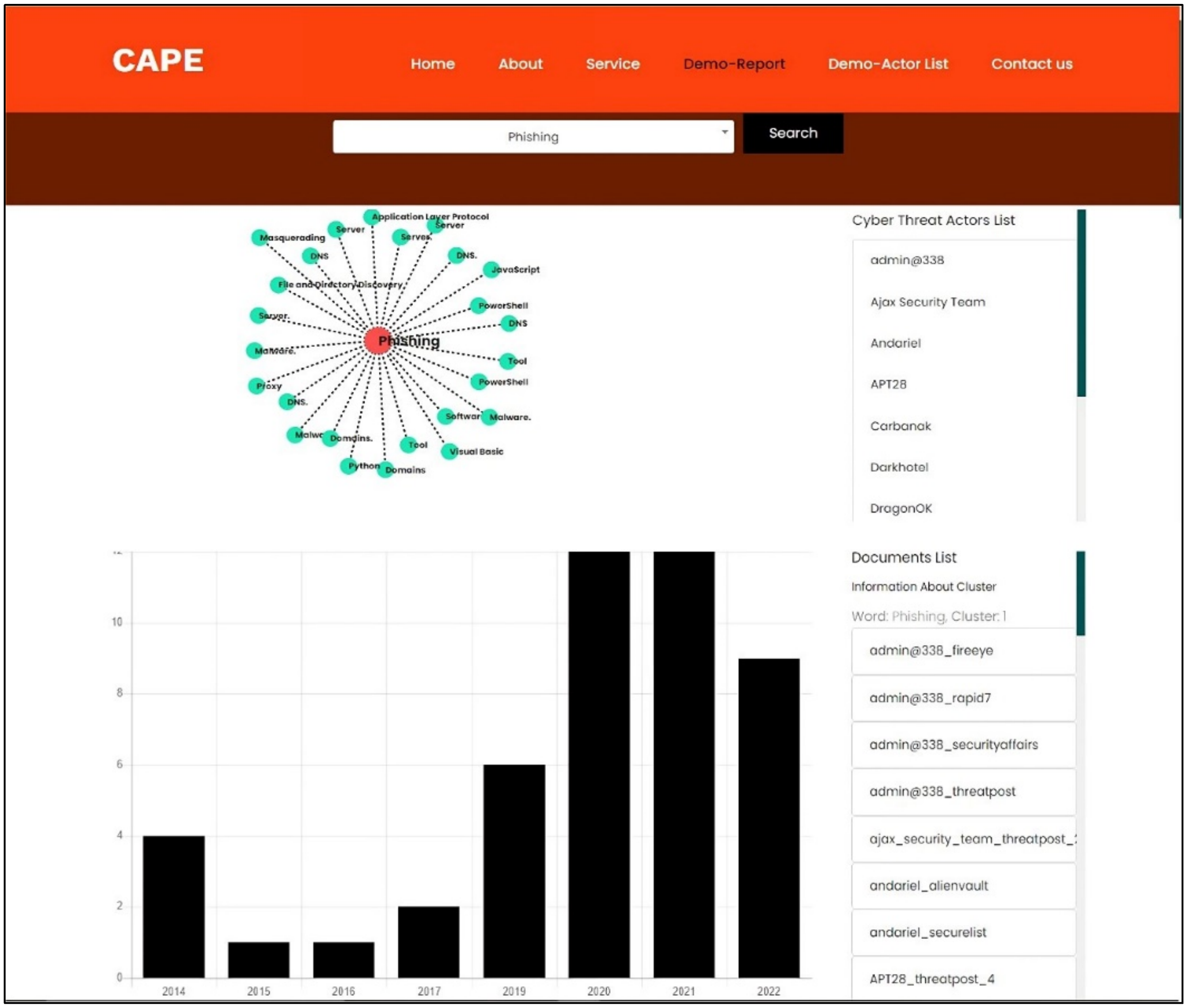

**Figure 6 -** User Interface of Cyber Attack Pattern Explorer (CAEP)

Participants performed the following tasks to assess the precision rate in document retrieval:

1) Participants entered specific attack patterns of their choice in the search bar.
2) Participants ensured that the targeted attack patterns were selected, with the node of the chosen attack patterns highlighted in red.
3) Participants examined the list of documents retrieved by the system.
4) Participants assessed the relevance of documents to the selected attack patterns to evaluate the precision rate.
5) The process was repeated for the cyber threat actor list.

The precision rate is defined as the ratio of documents retrieved by the system that participants consider most relevant. Table 5 presents the precision rate results for textual CTI documents and the list of cyber threat actors. Additionally, Table 6 displays the participant responses to each question, along with the mean rating for the questionnaire.

**Table 4** - CAPE Evaluation Requirements, Visual Components, and Evaluation methods

| | Requirement | Visual components | Evaluation methods |
|---|---|---|---|
| **Simplicity** | G1: Documents collection | 1. Document list<br>2. Cyber threat actor list<br>3. Documents list and view | Precision rate |
| **Clarity** | G2: Relevant attack patterns<br>G3: Cyber-attack patterns evolution | 1. Graph view<br>2. Timeline-based graph | Usability questionnaires |

The rating scores range from 1 to 5, with 5 indicating the highest rating. The scores are distributed within this range, with higher values being preferable. The overall average score is 4.0, with four scores exceeding 4.0, and the remaining four surpassing 3.4. While the majority of participants offered valuable feedback on usability, there is an evident need to improve performance and results.

**Table 5** - Precision Rates for Assessed Attack Patterns

| Evaluated attack patterns | Precision rate (Textual CTI Documents %) | Precision rate (Cyber threat actor list %) |
|---|---|---|
| External Proxy | 72 | 78 |
| Malicious File | 65.5 | 63 |
| Email Account | 95 | 90 |
| Malware | 100 | 100 |
| Mavinject | 89 | 83 |
| Remote Desktop Protocol | 91 | 95 |
| Non-Application Layer Protocol | 96 | 98 |
| Asymmetric Cryptography | 97 | 100 |
| Trap | 91 | 97 |
| Software | 86 | 84 |
| **Average** | **88.25** | **88.8** |

**Table 6** - Usability Assessment of the CAPE

| Questions | 1 | 2 | 3 | 4 | 5 | Average |
|---|---|---|---|---|---|---|
| Software performs the intended tasks | | | | 5 | 5 | 4.5 |
| Functionalities involved in the system are sufficient | | | | 6 | 4 | 4.4 |
| System is able to give expected results | | | | 5 | 5 | 4.5 |
| System interacts quickly | | | | 4 | 6 | 4.6 |
| I can comprehend and learn to use the system easily | | | 2 | 3 | 5 | 4.3 |
| Interface looks good and provides all required information | | | | 4 | 6 | 4.6 |
| Usage of the system is intuitive | | | 1 | 6 | 3 | 4.2 |
| Software is capable of handling errors | | | | 5 | 5 | 3.5 |

## 4.2 Performance analysis

The evaluation of the proposed work is presented in this section, focusing on assessing the capability of non-parametric approach to extract attack patterns form unstructured CTI documents. The performance analysis utilizes accuracy, F1-score, recall, precision, and false positive rate as evaluation parameters. Five machine learning algorithms are trained on the attack patterns dataset to gauge the effectiveness of the non-parametric approach (Section 3.6). The performance of the proposed approach is compared with LSA-based search system [5]. Furthermore, the results derived from the attack pattern dataset are compared with those from the LSA-based search system and the MITRE ATT&CK dataset [38].

**Table 7** - Empirical Evaluation of Proposed Cyber Threat Attribution Approach

| ML Algorithm | Dataset | Accuracy (%) | Recall (%) | Precision (%) | F1-score | False Positive Rate (%) |
|---|---|---|---|---|---|---|
| **Random Forest** | Non-Parametric Approach | 91.05 | 91.05 | 86.90 | 0.88 | 3 |
| | LSA | 88.00 | 89.00 | 92.00 | 0.89 | 3 |
| | ATT&CK | 74.00 | 74.00 | 65.67 | 0.7 | 6 |
| **Naïve Bayes** | Non-Parametric Approach | 95.32 | 95.56 | 91.91 | 0.93 | 2 |
| | LSA | 88.00 | 82.00 | 89.00 | 0.83 | 3 |
| | ATT&CK | 98 .00 | 98.00 | 97.00 | 0.97 | 1 |
| **Decision Tree** | Non-Parametric Approach | 77.69 | 72.64 | 75.17 | 0.73 | 4 |
| | LSA | 82.00 | 74.00 | 76.00 | 0.73 | 4 |
| | ATT&CK | 36.00 | 36.00 | 13.73 | 0.2 | 14 |
| **Deep Learning** | Non-Parametric Approach | 91.74 | 89.38 | 96.20 | 0.92 | 2 |
| | LSA | 94.00 | 89.00 | 90.00 | 0.89 | 3 |
| | ATT&CK | 70.00 | 70.00 | 61.83 | 0.66 | 6 |
| **K-NN** | Non-Parametric Approach | 80.72 | 76.77 | 94.20 | 0.84 | 2 |
| | LSA | 68.00 | 70.00 | 69.00 | 0.67 | 7 |
| | ATT&CK | 42.00 | 42.00 | 33.80 | 0.37 | 4 |

To evaluate the performance of the five machine learning algorithms, a cross-validation approach is utilized. This method involves splitting the original dataset into two segments: one for training the machine learning model and the other for evaluating its performance. Specifically, the dataset is dynamically and uniformly partitioned into k subgroups for k-fold cross-validation. In each iteration of this process, one of the k subsets serves as the test data, while the remaining k-1 subsets used as training data. This process is repeated k times, allowing each subgroup to take a turn as the test data. By averaging the outcomes across these iterations, an approximate assessment of the algorithms' performance is obtained. One significant benefit of this technique is that each instance has an equal chance of being used for evaluation and training. For this experiment, the value of k is set to 10. High-level IOCs that incorporate the attack patterns are utilized as the representation of each cyber threat actor's features. Subsequently, machine learning algorithms are applied to the dataset. All evaluation results are described in Table 7, which illustrates accuracy, precision, recall, f1-score, and false positive rate for three datasets.

Among these datasets, two are generated using the non-parametric approach and LSA while one is manually created from MITRE ATT&CK website. This table illustrates that NB achieves high accuracy (95.32%), recall (95.56%), and F1-score (0.93). Conversely, DLNN achieves a high precision of 96.20%. These results highlight the effectiveness of NB, DLNN, and RF in terms of higher accuracy, recall, precision, and F1-score along with low false positive rates. In contrast, DT and KNN perform poorly on the dataset generated by non-parametric approach. For the LSA-generated dataset, DLNN achieves high accuracy of 94%, while RF attains a high precision of 92%. The recall and F1-score for both algorithms are equal, at 89% and 0.89 respectively. Additionally, the MITRE ATT&CK dataset is utilized to train five machine learning algorithms comprising single-instance attack patterns against each cyber threat actor.

Consequently, the attribution of cyber incidents is not evaluated using cross-validation approach. To assess the performance of the machine learning algorithms with the MITRE ATT&CK dataset, a separate collection of test data is generated from recent incidents associated with cyber threat actors. The results indicate that NB outperforms for ATT&CK dataset, achieving high accuracy, recall, precision, F1-score, and false positive rate, which are 98%, 98%, 97%, 0.97, and 1% respectively. However, the results of other algorithms for ATT&CK dataset are comparatively lower than those for the other two datasets. The accuracy, recall, precision, F1-score, and false positive rate of each algorithm for the three datasets are compared separately in Figure 7 to Figure 11.

Figure 7 illustrates the accuracy of the RF algorithm across three datasets: non-parametric approach, LSA, and ATT&CK at 91.05%, 88%, and 74%, respectively. Notably, the non-parametric approach demonstrates highest accuracy with the RF algorithm. The accuracy of NB for the three datasets is 95.32%, 88%, and 98%, highlighting MITRE ATT&CK's effectiveness with this algorithm. However, as mentioned earlier, the ATT&CK dataset has limitations due to manual maintenance, leading to unreported attack patterns. The accuracy of DT for the three datasets is 77.69%, 82%, and 36%, emphasizing LSA's superiority over the non-parametric approach and ATT&CK. In deep learning, LSA achieves 94% accuracy, outperforming non-parametric (91.74%) and ATT&CK (70%). Regarding KNN, the non-parametric approach attains high accuracy (80.72%) compared to LSA and ATT&CK, where accuracy is 68% and 42%, respectively.

Figure 8 depicts recall values for RF across three datasets (non-parametric approach, LSA, and ATT&CK) at 91.05%, 89%, and 74%, respectively. The non-parametric approach consistently exhibits higher recall than the other methods. NB yields recall rates of 95.56%, 82%, and 98%, highlighting ATT&CK's effectiveness with this algorithm. DT recall for the three datasets is 72.64%, 74%, and 36%, emphasizing LSA's superior performance over the non-parametric approach and ATT&CK. In DLNN, the non-parametric approach achieves high recall of 89.38%, surpassing LSA (89%) and ATT&CK (70%). Using KNN, the non-parametric approach achieves high recall (76.77%) compared to LSA and ATT&CK, where recall is 70% and 42%, respectively. This comparison highlights the non-parametric approach's strong performance in recall across RF, DLNN, and KNN.

In Figure 9, precision values for RF across the three datasets (non-parametric approach, LSA, and ATT&CK) are 86.90%, 92%, and 65.67%, respectively. Notably, LSA demonstrates the highest precision compared to the non-parametric approach and ATT&CK. NB precision for the three datasets is 91.91%, 89%, and 97%, highlighting ATT&CK's effectiveness with this algorithm. In the case of LSA using decision tree, precision is 76%, surpassing non-parametric and ATT&CK (75.17% and 13.73%, respectively). For DLNN, precision is 96.20%, 90%, and 61.83% for the non-parametric approach, LSA, and ATT&CK, respectively, highlighting the non-parametric approach's superior performance. Using KNN, the non-parametric approach achieves the highest precision at 94.20%, while LSA and ATT&CK have precisions of 69% and 33.80%, respectively. Overall, the non-parametric approach excels in precision with both DLNN and KNN.

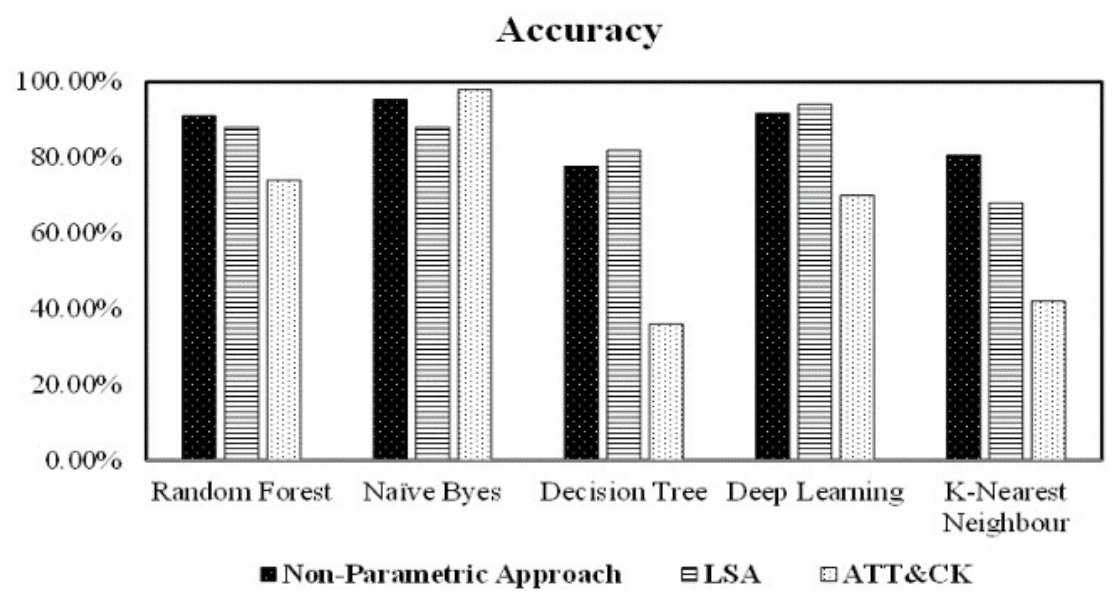

**Figure 7** - Accuracy

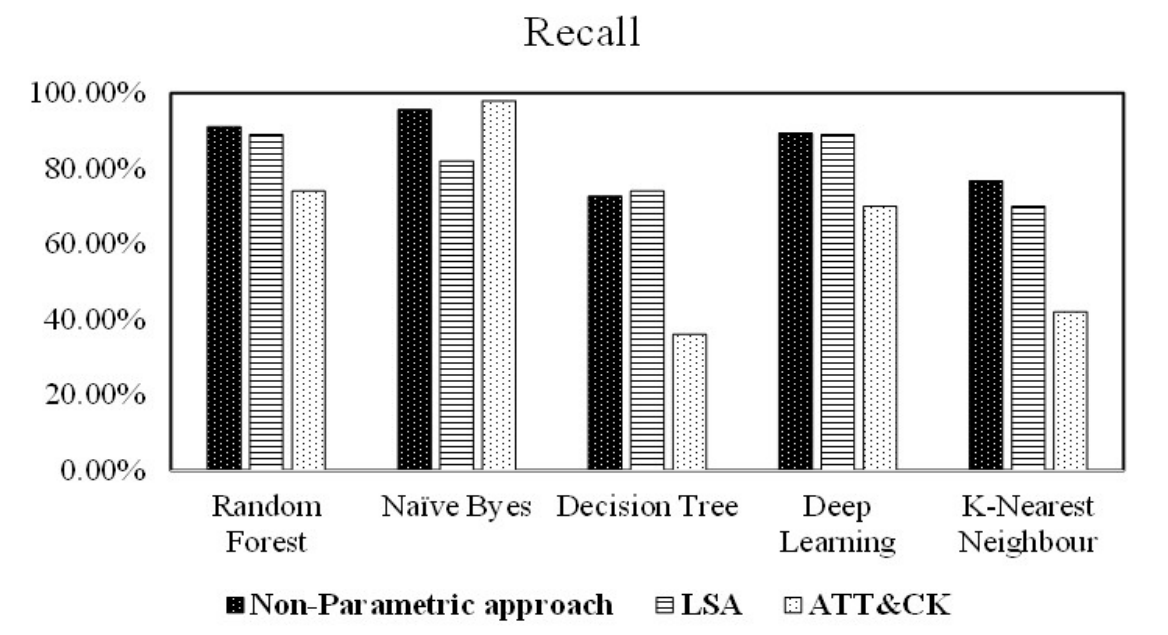

**Figure 8 - Recall**

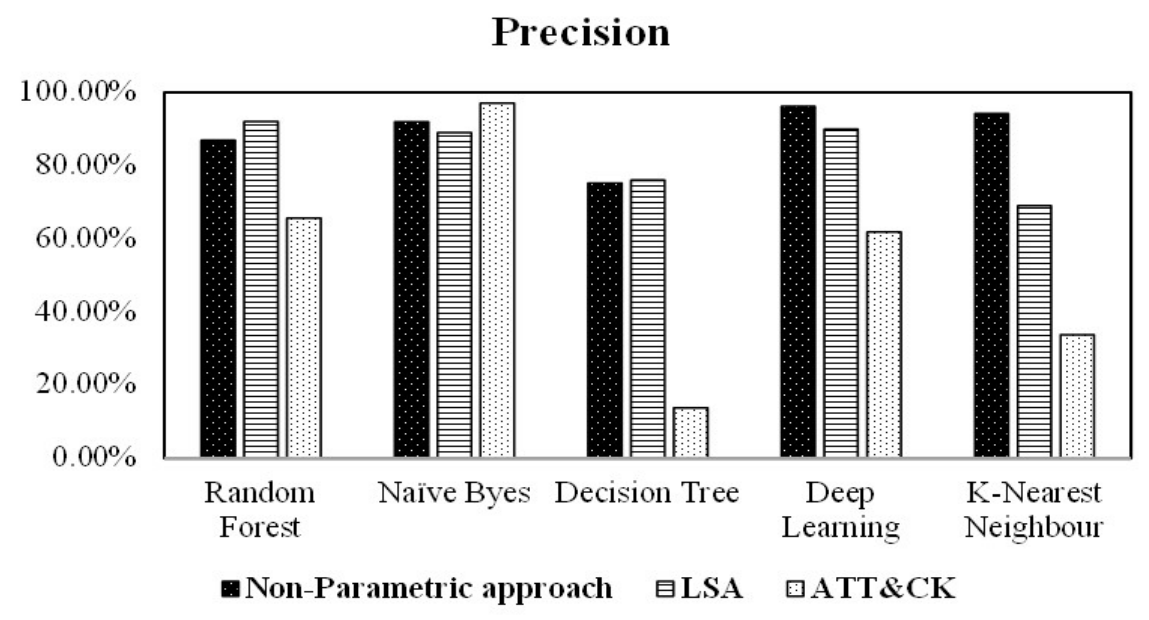

**Figure 9 -** Precision

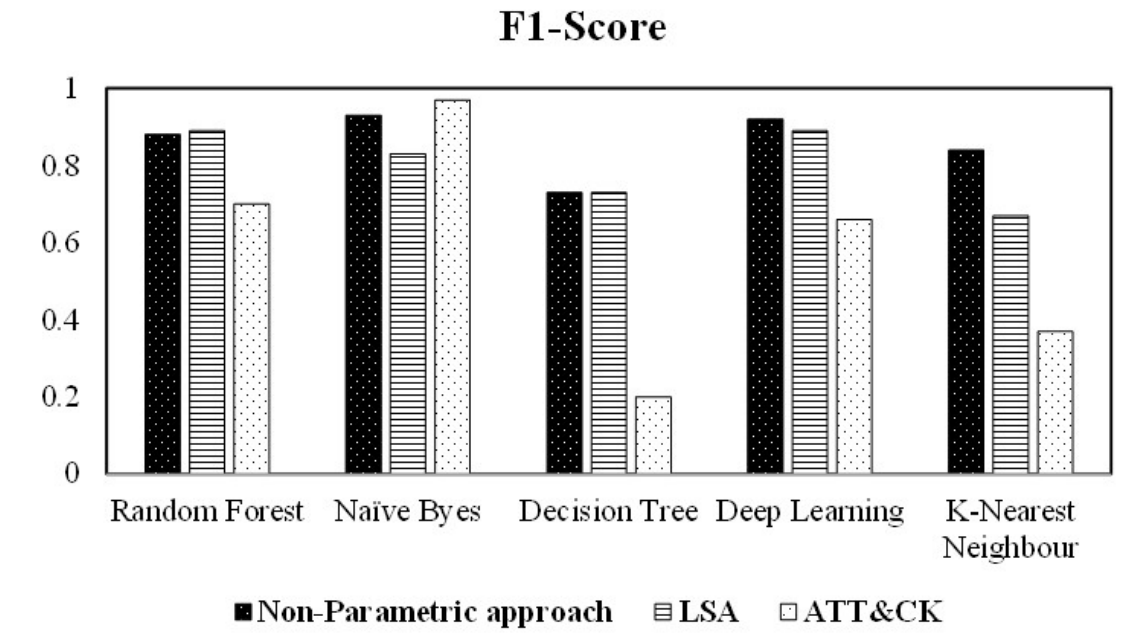

**Figure 10 -** F1 Scores

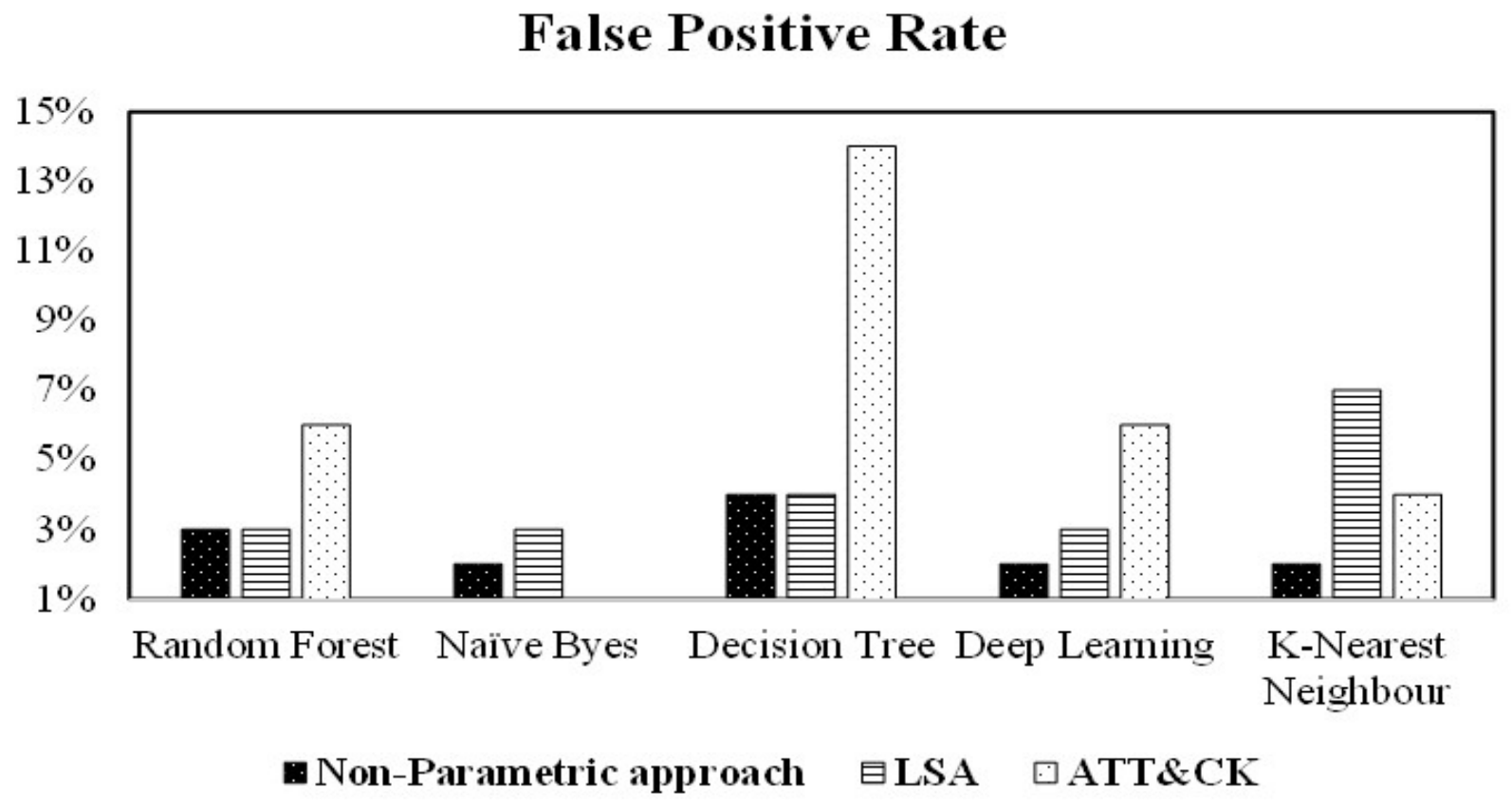

**Figure 11 -** False Positive Rate

In Figure 10, LSA achieves a high F1-score of 0.89 using RF, surpassing the F1-scores of non-parametric and ATT&CK, which are 0.88 and 0.70, respectively. The F1-score using NB for the three datasets is 0.93, 0.83, and 0.97, indicating that the MITRE ATT&CK dataset attains the highest F1-score compared to non-parametric and LSA. Non-parametric and LSA achieve an equal F1-score of 0.73 using DT, while ATT&CK scores 0.2. For DLNN, the non-parametric approach attains an F1-score of 0.92, outperforming LSA and ATT&CK, with F1-scores of 0.89 and 0.66, respectively. In the case of KNN, the non-parametric approach achieves F1-scores of 0.84, surpassing LSA (0.67) and ATT&CK (0.37). Notably, the non-parametric approach consistently achieves high F1-scores across RF, DLNN, and KNN.

In Figure 11**,** it is shown that the false positive rate using RF for the non-parametric approach and LSA is equal (i.e., 3%) while the false positive rate for ATT&CK is 6%. Using NB, ATT&CK achieve low false positive rate of 1%, while the false positive rate of non-parametric approach and LSA is 2% and 3% respectively. The false positive rate using DT for the three datasets (i.e., non-parametric approach, LSA, and ATT&CK) is 4%, 4%, and 14% respectively, indicating that ATT&CK has a high false positive rate using DT algorithm. The non-parametric approach has a low false positive rate using DLNN, which is 2% as compared to LSA and ATT&CK. The false positive rate of LSA and ATT&CK is 3% and 6% respectively. The false positive rate using KNN for non-parametric approach is 2% which is low as compared to others. The false positive rate using KNN for LSA and ATT&CK is 7% and 4% respectively. It is clear that non-parametric approach has low false positive rate using RF, DT, DLNN, and KNN. In the following section, the outcomes of these machine learning algorithms are evaluated for every unique cyber threat actor.

Figure 12 to Figure 16 present precision comparisons for individual cyber threat actors across the three datasets. The LSA-based dataset constructed by Noor et al. is limited to thirty six cyber threat actors [5], while in this study, we consider fifty threat actors. In Figure 12, precision using RF reveal challenges in identifying thirteen cyber threat actors, including: *Axiom, Black Oasis, Bronze Butler, Lotus Blossom, Night Dragon, Pitty Tiger, Putter Panda, Silent Librarian, Suckfly, TA459, Thrip, Whitefly,* and *Winnti Group*. Some threat actors, such as *Carbanak, DragonOK*, and *Gorgon Group*, are identified with low precision. The non-parametric approach successfully identifies all cyber threat actors using RF, although four actors (*i.e., Night Dragon, Orange Worm, TA459, and Whitefly*) exhibit low precision. In the case of LSA, RF falls short in identifying five cyber threat actors (*i.e., DragonOk, Equation, Gorgon Group, Magic Hound, and Pitty Tiger*).

Figure 13 reveals that NB fails to identify the cyber threat actor *Lotus Blossom* in the ATT&CK dataset, and *DragonOk* is identified with low precision. In the case of LSA, three cyber threat actors, namely *APT41, Andariel, and Blue Mockingbird,* are identified with low precision. Similarly, using the non-parametric approach, three cyber threat actors, namely *Pitty Tiger, Thrip, and Whitefly* are identified with low precision.

Furthermore, DLNN struggles to identify fifteen cyber threat actors in the ATT&CK dataset, including: *BlackOasis, CostaRicto, DragonOK, ElderWood, Ferocious Kitten, Frankenstein, Gallium, Lotus Blossom, Naikon, Orangeworm, Pitty Tiger, Suckfly, Thrip, Whitefly, and Zirconium. Additionally, Admin@338, Axiom, FIN10, Rancor, TA459,* and *Winnti Group* are identified with low precision using DLNN for the ATT&CK dataset.

In the non-parametric approach, DLNN successfully identifies all cyber threat actors, but *Lotus Blossom, Magic Hound,* and *Putter Panda* are identified with low precision. For LSA, *Equation* is not identified using DLNN, and *Magic Hound* is identified with low precision in the LSA dataset.

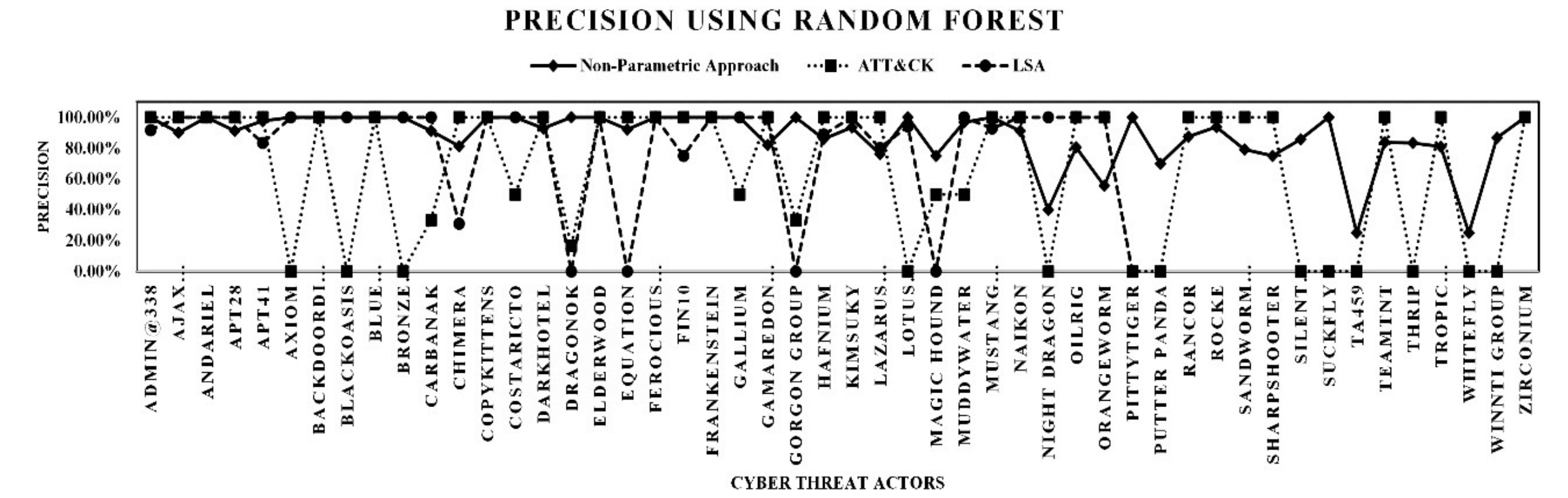

**Figure 12**. Precision Comparison of Individual Cyber Threat Actors for the Three Datasets Using RF

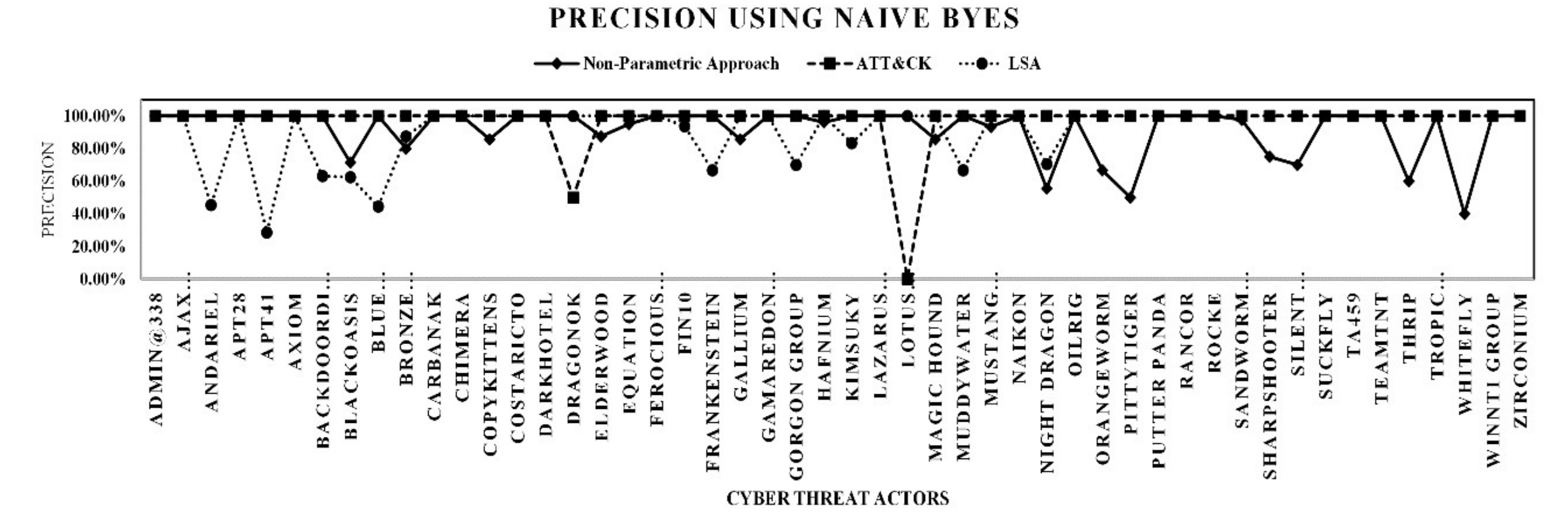

**Figure 13**. Precision Comparison of Individual Cyber Threat Actors for the Three Datasets Using NB

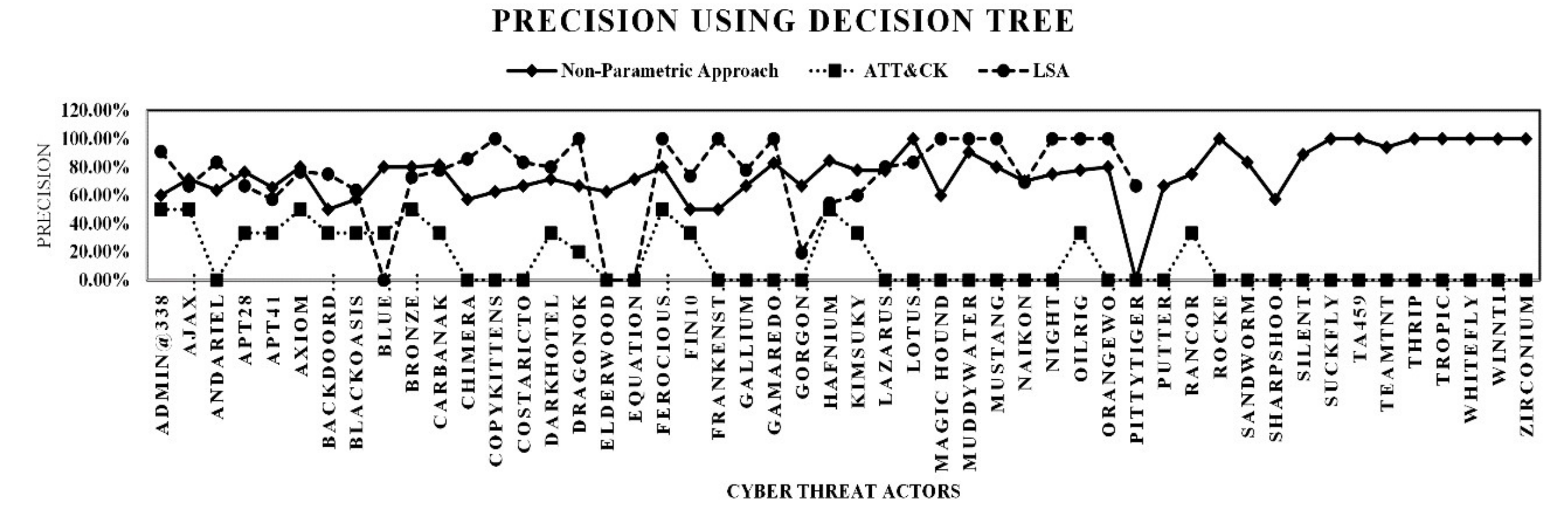

**Figure 14**. Precision Comparison of Individual Cyber Threat Actors for the Three Datasets Using DT

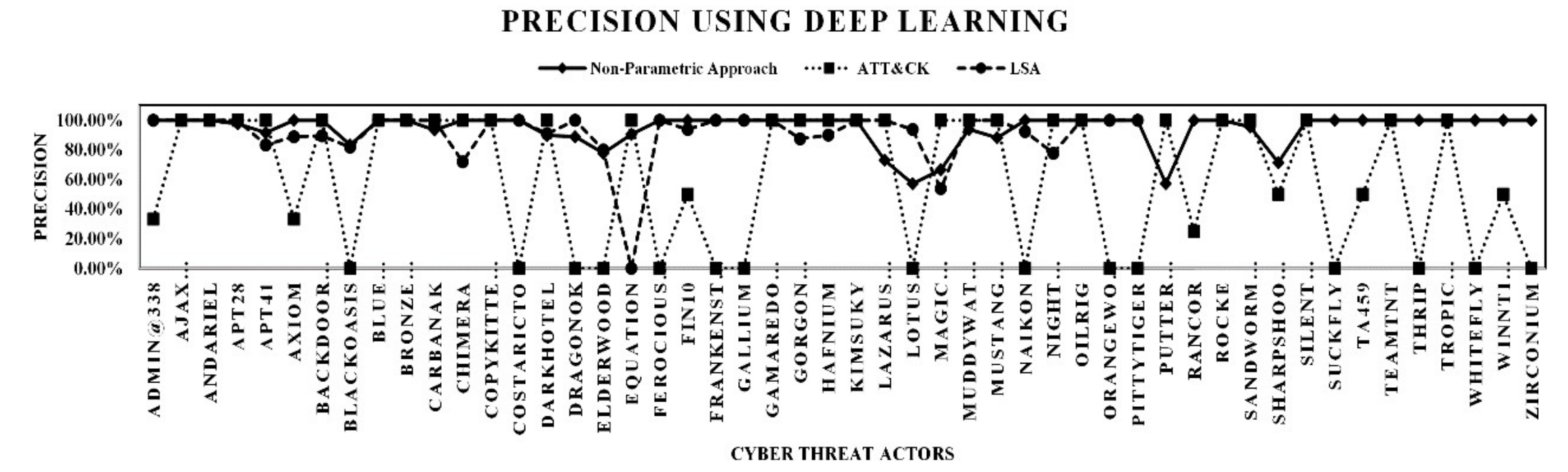

**Figure 15**. Precision Comparison of Individual Cyber Threat Actors for the Three Datasets Using DLNN

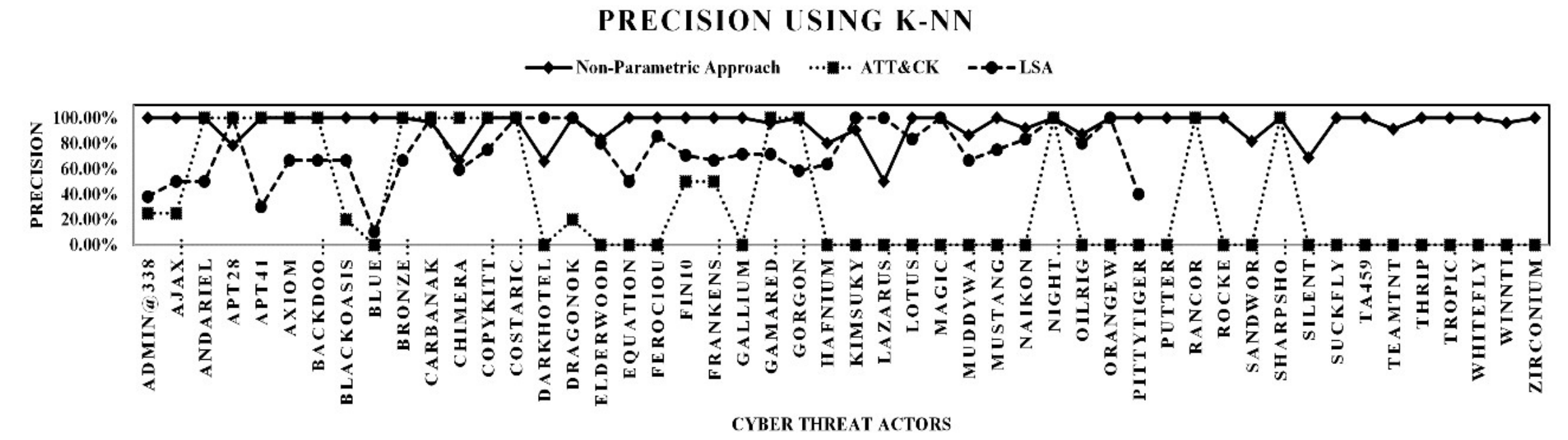

**Figure 16**. Precision Comparison of Individual Cyber Threat Actors for the Three Datasets Using KNN

In Figure 14, it is evident that the DT struggles to identify thirty-two cyber threat actors in the ATT&CK dataset, including: *Andarial, Chimera, CopyKittens, CostaRicto, Elderwood, Equation, Frankenstien, Gallium, Gamaredon Group, Gorgon Group, Lazarus Group, Lotus Blossom, Magic Hound, Muddy Water, Mustang Panda, Naikon, Night Dragon, Orangeworm, Pitty Tiger, Putter Panda, Rocke, Sandworm Team, Sharoshooter, Silent Librarian, Suckfly, TA459, TeamTNT, Thrip, Tropic Trooper, Whitefly, Winnti Group,* and *Zirconium*. The remaining cyber threat actors are identified with low precision using DT in the ATT&CK dataset. For LSA, DT fails to identify three cyber threat actors, namely *Blue Mockingbird, Elderwood,* and *Equation* and identifies three other actors with low precision. In the case of the non-parametric approach, DT cannot identify one cyber threat actor *Pitty Tiger* resulting in the identification of a few other actors with low precision.

In Figure 15, DLNN struggles to identify fifteen cyber threat actors in the ATT&CK dataset, including: *BlackOasis, CostaRicto, DragonOK, ElderWood, Ferocious Kitten, Frankenstein, Gallium, Lotus Blossom, Naikon, Orangeworm, Pitty Tiger, Suckfly, Thrip, Whitefly, and Zirconium. Additionally, Admin@338, Axiom, FIN10, Rancor, TA459,* and *Winnti Group* are identified with low precision using DLNN for the ATT&CK dataset. In the non-parametric approach, DLNN successfully identifies all cyber threat actors, but *Lotus Blossom, Magic Hound,* and *Putter Panda* are identified with low precision. For LSA, *Equation* is not identified using DLNN, and *Magic Hound* is identified with low precision in the LSA dataset.

In Figure 16**,** KNN encounters challenges in identifying twenty-nine cyber threat actors in the ATT&CK dataset, including: *Blue Mockingbird, Darkhotel, Elderwood, Equation, Ferocious Kitten, Gallium, Hafnium, Kimsuky, Lazarus Group, Lotus Blossom, Magic Hound, MuddyWater, Mustang Panda, Naikon,*

*Oilrig, Orangeworm, Pitty Tiger, Putter Panda, Rocke, Sandworm Team, Silent Librarian, Suckfly, TA459, TeamTNT, Thrip, Tropic Trooper, Whitefly, Winnti Group,* and *Zirconium*. Additionally, *Admin@338, Ajax Security Team, BlackOasis, DragonOK, FIN10,* and *Frankenstein* are identified with low precision using KNN for the ATT&CK dataset. In the case of LSA, all cyber threat actors are successfully identified, although *Admin@338, APT41, Blue Mockingbird,* and *Pitty Tiger* are identified with low precision. For the non-parametric approach, a few cyber threat actors namely *Chimera, Darkhotel,* and *Lazarus Group* are identified with low precision.

In conclusion, the analysis indicates that NB demonstrates superior precision rates for the ATT&CK dataset. Within the non-parametric approach, DLNN exhibits commendable precision rate. Meanwhile, for LSA, the precision rate is notably high when utilizing the RF algorithm. These findings highlight the importance of selecting appropriate algorithms tailored to specific datasets to optimize precision in cyber threat actor identification. Overall, the precision of the non-parametric approach-based dataset is better comparing to those based on LSA and the ATT&CK dataset for all fifty cyber threat actors.

In continuation, Figure 17 to Figure 21 depict the recall rates for individual cyber threat actors across three datasets. The LSA-based dataset is limited to 36 cyber threat actors. In Figure 17, the recall rate of RF is explored for the non-parametric approach, LSA, and ATT&CK datasets. Notably, RF demonstrates the lowest recall for thirteen cyber threat actors: *Axiom, BlackOasis, BRONZE BUTLER, Lotus Blossom, Night Dragon, Pitty Tiger, Putter Panda, Silent Librarian, Suckfly, TA459, Thrip, Whitefly,* and *Winnti Group*. The Figure 17 also reveals that, for LSA, RF has the lowest recall for *DragonOK, Equation, Gorgon Group, Magic Hound,* and *Pitty Tiger*. In the case of the non-parametric approach *TA459* is identified with low recall using RF algorithm.

In Figure 18**,** it is evident that NB exhibits the lowest recall for one cyber threat actor *Lotus Blossom* in the ATT&CK dataset. Additionally, NB demonstrates low recall for five cyber threat actors in the LSA dataset: *Axisom, BlackOasis, Carbanak, Lotus Blossom* and *Orangeworm*. For the non-parametric approach, NB shows low recall for five cyber threat actors: *Backdoor Diplomacy, Chimera, Kimsuky, Mustang Panda,* and *Zirconium*.

Figure 19 depicts that DT exhibits the lowest recall for thirty-two cyber threat actors in the ATT&CK dataset: *Andariel, Chimera, CopyKittens, CostaRicto, Elderwood, Equation, Frankenstein, Gallium, Gamaredon, Gorgon Group, Lazarus Group, Lotus Blossom, Magic Hound, MuddyWater, Mustang Panda, Naikon, Night Dragon, Orangeworm, Pitty Tiger, Putter Panda, Rocke, Sandworm Team, Sharpshooter, Silent Librarian, Suckfly, TA459, Team TNT, Thrip, Tropic Trooper, Whitefly, Winnti Group,* and *Zirconium*. For LSA, DT has the lowest recall for *Blue Mockingbird, Elderwood, Equation, Kimsuky, Magic Hound,* and *Orangeworm*. In the non-parametric approach, DT exhibits the lowest recall for fifteen cyber threat actors namely *Backdoor Diplomacy, Ferocious Kitten, FIN10, Frankenstein, Gorgon Group, Magic Hound, Pitty Tiger, Putter Panda, Rocke, Silent Librarian, Tropic Trooper, Whitefly, Winnti Group,* and *Zirconium*.

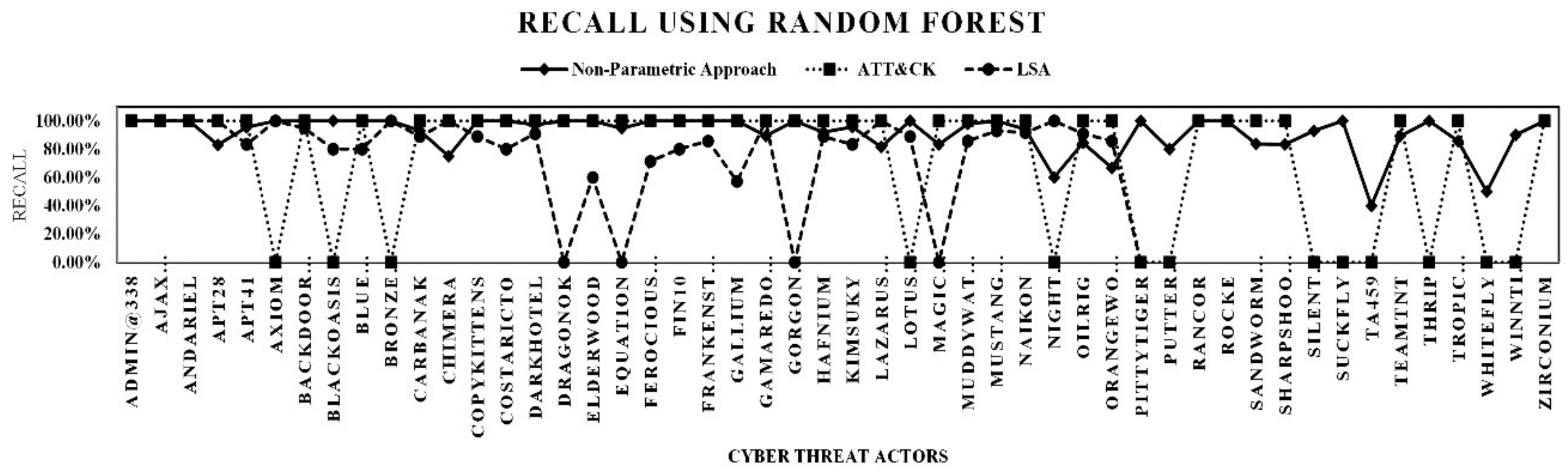

**Figure 17**. Recall Comparison of Individual Cyber Threat Actors for the Three Datasets Using RF

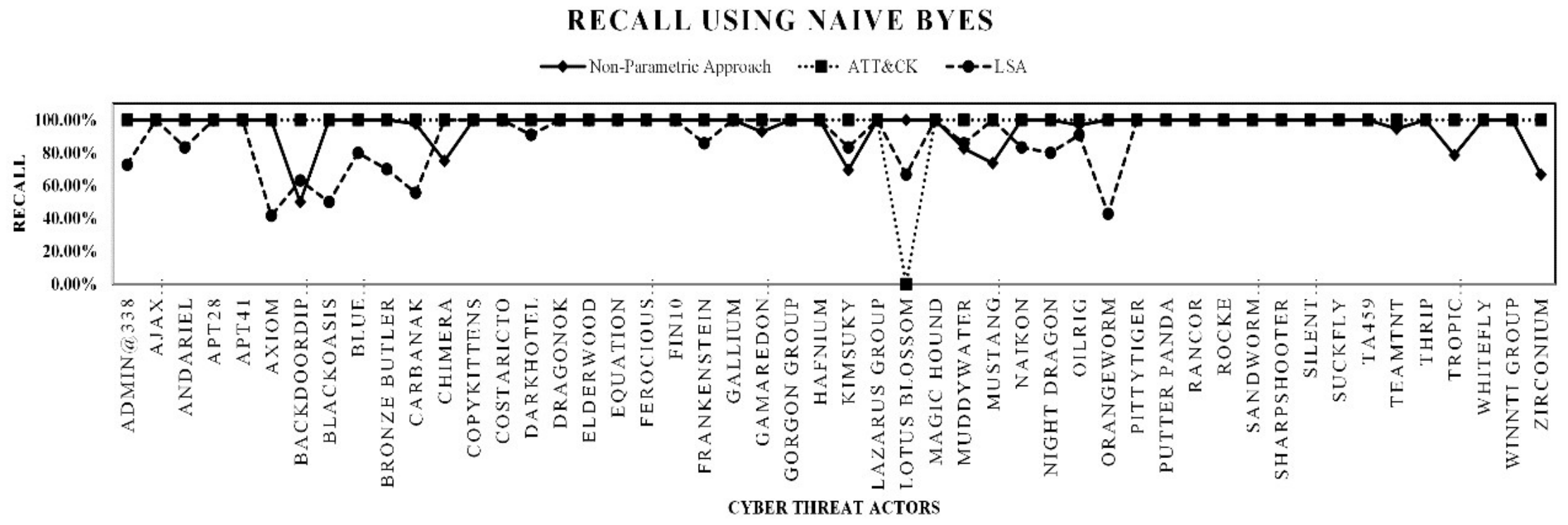

**Figure 18**. Recall Comparison of Individual Cyber Threat Actors for the Three Datasets Using NB

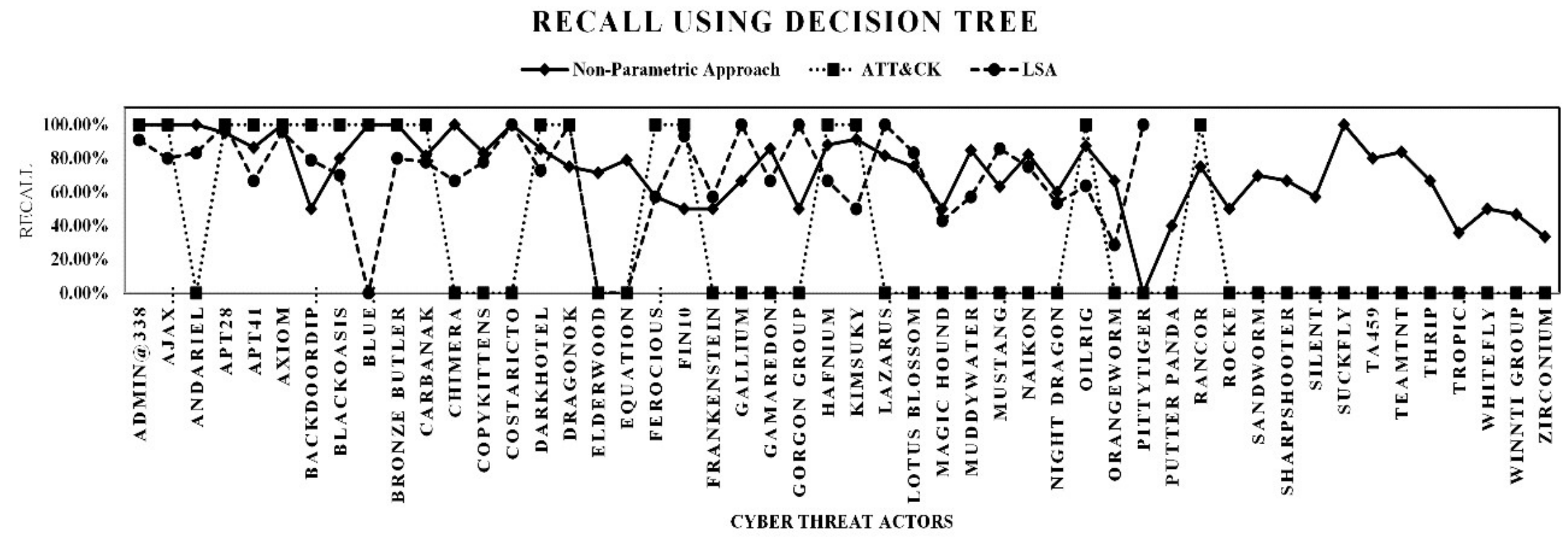

**Figure 19**. Recall Comparison of Individual Cyber Threat Actors for the Three Datasets Using DT

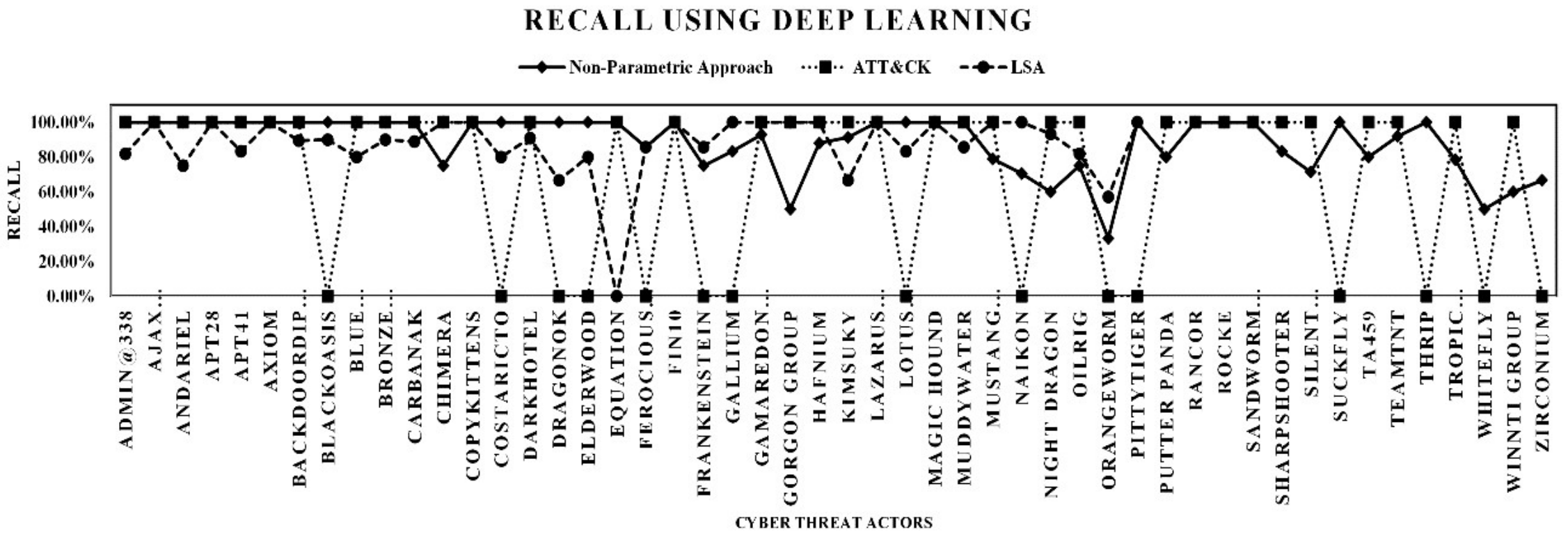

**Figure 20**. Recall Comparison of Individual Cyber Threat Actors for the Three Datasets Using DLNN

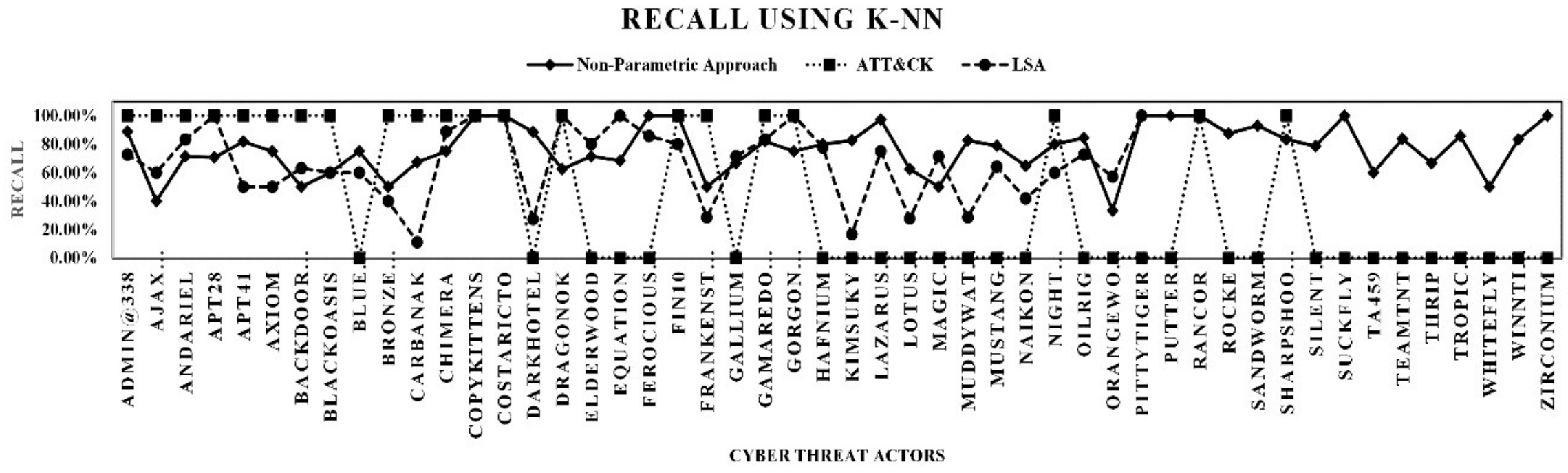

**Figure 21**. Recall Comparison of Individual Cyber Threat Actors for the Three Datasets Using KNN

In Figure 20**,** it is evident that DLNN exhibits the lowest recall for fifteen cyber threat actors in the ATT&CK dataset: *BlackOasis, CostaRicto, DragonOk, Elderwood, Ferocious Kitten, Frankenstein, Gallium, Lotus Blossom, Naikon, Orangeworm, Pitty Tiger, Suckfly, Thrip, Whitefly,* and *Zirconium*. For LSA, DLNN demonstrates low recall for four cyber threat actors, namely *DragonOk, Equation, Kimsuky,* and *Orangeworm*. In the non-parametric approach, DLNN shows low recall for *Gorgon Group, Orangeworm, Night Dragon, Winnti Group, Whitefly,* and *Zirconium*.

The Figure 21 illustrates that KNN exhibits the lowest recall for thirty cyber threat actors in the ATT&CK dataset, including *Blue Mockingbird, Darkhotel, Elderwood, Equation, Ferocious Kitten, Gallium, HAFINUM, Kimsuky, Lazarus Group, Lotus Blossom, Magic Hound, Muddywater, Mustang Panda, Naikon, Oilrig, Orangeworm, Pitty Tiger, Putter Panda, Rocke, Sandworm Team, Silent Librarian, Suckyfly, TA459, Team TNT, Thrip, Whitefly, Tropic Trooper, Winnti Group,* and *Zirconium*.

In the case of LSA, KNN demonstrates low recall for fourteen cyber threat actors, including *Ajax Security Team, APT41, Axiom, BlackOasis, Blue Mockingbird, Bronze Butler, Carbanak, Frankenstein, Kimsuky, Lotus Blossom, Muddywater, Mustang Panda, Naikon,* and *Orangeworm*. Additionally, KNN exhibits low recall for *Ajax Security Team, BackdoorDiplomacy, Bronze Butler, DragonOK, Frankenstein, Magic Hound, Orangeworm, TA459,* and *Whitefly*.

In conclusion, NB outperforms for the ATT&CK dataset and the non-parametric approach, while DLNN demonstrates high recall for LSA. Overall, the individual results for cyber threat actors indicate that the non-parametric approach, based on its dataset, exhibits better recall compared to both the LSA-based dataset and the ATT&CK dataset. This observation is made while considering the limitations associated with the latter two datasets.

The Figure 22 to Figure 26 present the F1-score for five machine learning models for each individual cyber threat actor. In Figure 22, it is observed that RF exhibits the lowest F1-score for fifteen cyber threat actors in the ATT&CK dataset, including *Axiom, BlackOasis, Bronze Butler, DragonOK, Lotus Blossom, Night Dragon, Pitty Tiger, Putter Panda, Silent Librarian, Suckfly, TA459, Thrip, Whitefly, and Winnti Group. For LSA, DragonOK, Equation, Gorgon Group,* and *Pitty Tiger* have low F1-scores. In the non-parametric approach, RF shows low F1-scores for four cyber threat actors, namely *Night Dragon, Orangeworm, TA459,* and *Whitefly*.

In Figure 23, it is evident that NB exhibits the lowest F1-score for one cyber threat actor, *Lotus Blossom*, in the ATT&CK dataset. For LSA, NB demonstrates low F1-scores for seven cyber threat actors: *Andariel, APT41, Axiom, Backdoor Diplomacy, BlackOasis, Blue Mockingbird,* and *Orangeworm*. In the non-parametric approach, NB shows low F1-scores for *Pitty Tiger* and *Whitefly*.

The Figure 24 reveals that DT exhibits the lowest F1-score for thirty-two cyber threat actors in the ATT&CK dataset, including *Andariel, Chimera, CopyKittens, CostaRicto, Elderwood, Equation, Frankenstein, Gallium, Gamaredon Group, Gorgon Group, Lazarus Group, Lotus Blossom, Magic Hound, Muddywater, Mustang Panda, Naikon, Night Dragon, Orangeworm, Pitty Tiger, Putter Panda, Rocke, Sandworm Team, Sharpshooter, Silent Librarian, Suckfly, TA459, TeamTNT, Thrip, Tropic Trooper, Whitefly, Winnti Group,* and *Zirconium*. For LSA, DT has low F1-scores for *Blue Mockingbird, Elderwood, Equation,* and *Gorgon Group*. In the non-parametric approach, DT exhibits low F1-scores for eight cyber threat actors: *FIN10, Frankenstein, Gorgon Group, Magic Hound, Pitty Tiger, Putter Panda, Tropic Trooper,* and *Zirconium*.

In Figure 25, it is evident that DLNN exhibits the lowest F1-score for fourteen cyber threat actors in the ATT&CK dataset, including *BlackOasis, CostaRicto, DragonOK, Elderwood, Ferocious Kitten, Frankenstein, Gallium, Lotus Blossom, Naikon, Orangeworm, Pitty Tiger, Suckfly, Thrip,* and *Whitefly*. For LSA, DLNN has a low F1-score for one cyber threat actor, *Equation*. In the non-parametric approach, there is a low F1-score for *Orangeworm* using DLNN.

The Figure 26 illustrates that KNN exhibits the lowest F1-score for thirty-two cyber threat actors in the ATT&CK dataset, including *BlackOasis, Blue Mockingbird, Darkhotel, DragonOK, Elderwood, Equation, Ferocious Kitten, Gallium, Hafinum, Kimsuky, Lazarus Group, Lotus Blossom, Magic Hound, Muddywater, Mustang Panda, Naikon, Oilrig, Orangeworm, Pitty Tiger, Putter Panda, Rocke, Sandworm Team, Silent Librarian, Suckfly, TA459, TeamTNT, Thrip, Tropic Trooper, Whitefly, Winnti Group,* and *Zirconium*.

In the case of LSA, KNN has a low F1-score for *Admin@338, Ajax Security Team, Andariel, APT41, Blue Mockingbird, Bronze Butler, Carbanak, Darkhotel, Frankenstein, Kimsuky, Lotus Blossom,* and *Muddywater*. For the non-parametric approach, KNN exhibits a low F1-score for two cyber threat actors, *Ajax Security Team* and *Orangeworm*. In conclusion, NB outperforms for the ATT&CK dataset and the non-parametric approach in achieving a high F1-score. For LSA, DLNN achieves a high F1-score. In summary, the individual results for cyber threat actors suggest that the non-parametric approach, leveraging its dataset, demonstrates a higher F1-score compared to both the LSA-based dataset and the ATT&CK dataset. This observation considers the limitations associated with the latter two datasets, particularly in terms of missing attack patterns.

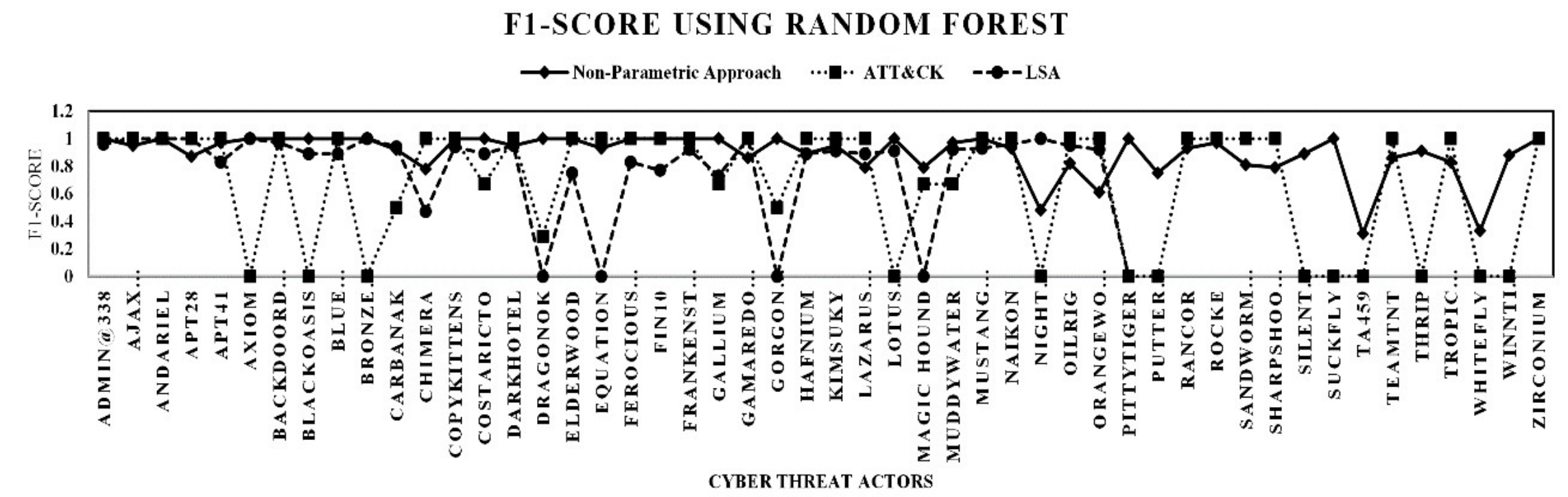

**Figure 22**. F1 Score Comparison of Individual Cyber Threat Actors for the Three Datasets Using Random Forest

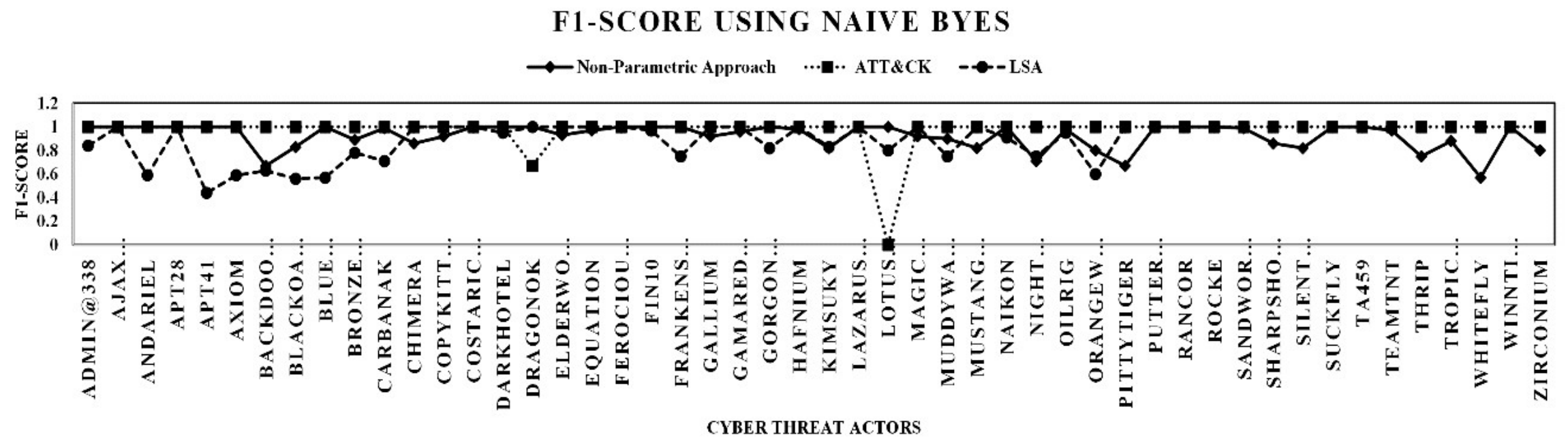

**Figure 23**. F1 Score Comparison of Individual Cyber Threat Actors for the Three Datasets Using Naive Bayes

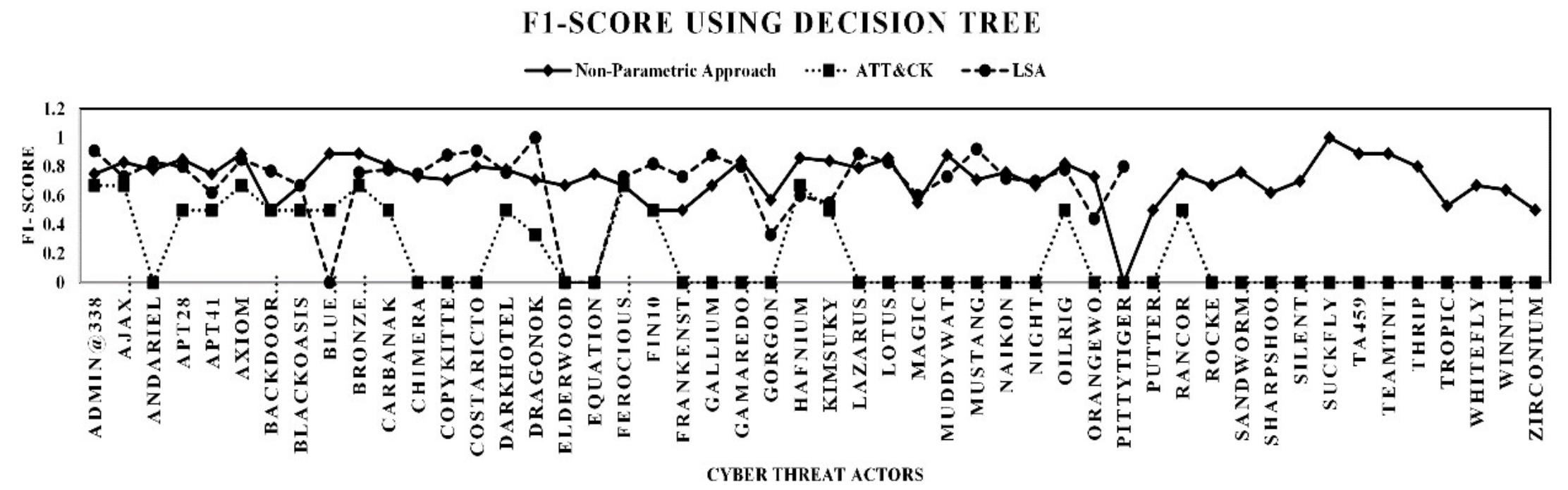

**Figure 24**. F1 Score Comparison of Individual Cyber Threat Actors for the Three Datasets Using Decision Tree

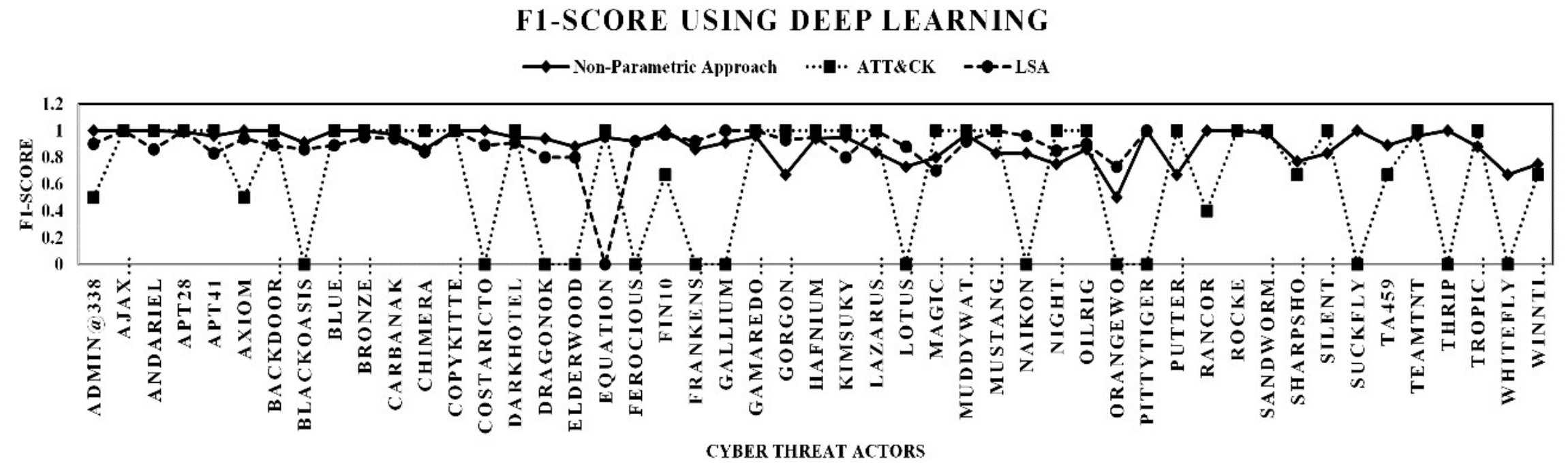

**Figure 25**. F1 Score Comparison of Individual Cyber Threat Actors for the Three Datasets Using Deep Learning

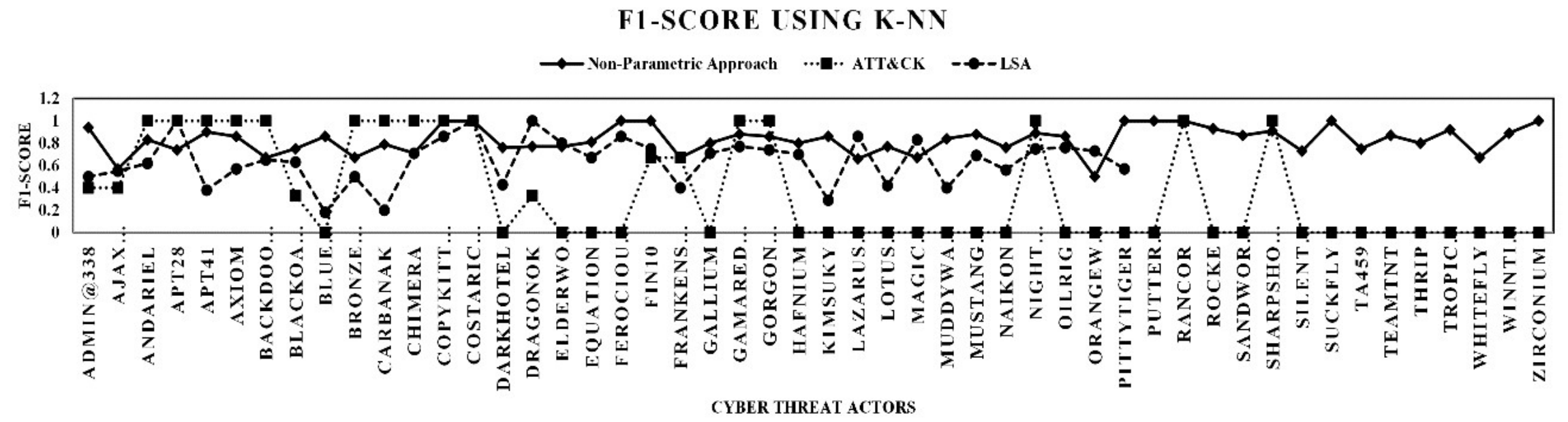

**Figure 26**. F1 Score Comparison of Individual Cyber Threat Actors for the Three Datasets Using K-NN

## 5. Discussion

The scenarios discussed in Section 5.1 illustrate how our approach, is assisted by the CAPE tool, can be applied in various real-world situations to enhance the effectiveness of cybersecurity analysts in identifying, analyzing, and responding to cyber threats. By providing an interactive and visually intuitive tool, CAPE empowers analysts to make informed decisions and improve their overall cybersecurity posture.

### 5.1 Real-world case descriptions

This section describes the real-world cases where CAPE tool can be utilized along with existing cybersecurity tools to effectively attribute cyber threat actors. CAPE is a visually interactive analytics tool designed to facilitate efficient information discovery through interactive visualization and mining techniques.

#### 5.1.1 Scenario: Incident Response and Threat Attribution

Consider an incident at a financial institution where a data breach was occurred, resulting in a considerable amount of data being exfiltrated. The cybersecurity team is investigating the incident, and initial indicators suggest that the breach involves sophisticated attack patterns previously observed in similar incidents.

The investigations revealed exploit attempts on public-facing applications (Exploit Public-Facing Application, T1190) [69] for gaining initial access. It is detected that the threat actor attempted to exploit against vulnerabilities like CVE-2017-1000486, CVE-2015-7450, and CVE-2010-5326 [69]. Additionally, unusual remote access activity (External Remote Services, T1133) [70] was observed. These remote access attempts utilized anomalous remote access services such as VPNs, particularly from unusual IP addresses or during non-standard working hours [70].

The cybersecurity team also observed traces of credential access and discovery during the investigation of data breach incident. They detected suspicious processes (OS Credential Dumping: LSASS Memory, T1003.001) [71] such as “ProcDump” and “Mimikatz” running on systems, which are used to dump credentials from memory. Furthermore, traces of unusual access to the registry (OS Credential Dumping: Security Account Manager, T1003.002) [72] or modifications to the SAM registry hive indicating attempts to extract password hashes are found during the investigations. The investigations also revealed traces of network scans using tools like nmap, especially targeting MS-SQL servers (Network Service Discovery, T1046) [73].

Regarding persistence and privilege escalation indicators, the cybersecurity team observed traces of registry modifications (Boot or Logon Autostart Execution: Registry Run Keys / Startup Folder, T1547.001) [74]. They also discovered new local or domain accounts, especially those with administrative privileges [74], and noted changes in user roles and privileges, including the granting of sysadmin roles to new accounts [75].

After gaining persistent access, it was observed that the threat actors performed the lateral movement and execution activities. Traces were found indicating the execution of PowerShell commands and scripts, particularly those querying DNS or executing remote commands (T1059.001) [76]. They also created the scheduled tasks that correspond to known malicious patterns (T1053.005) [77]. Furthermore, there are traces of unusual or unauthorized use of remote services like RDP, SMB, SSH, and WinRM for lateral movement (T1021.001, T1021.002, T1021.004, T1021.006) [78]–[81].

During investigations, the traces of data collection and exfiltration are observed. The threat actors used the temporary directories such as `C:\Windows\Temp` or `/tmp` for staging data (T1074.001) [82]. The threat actors used the compression tools like 7zip to archive data, especially in conjunction with other suspicious activities (T1560.001) [83]. There are unusual outbound HTTP requests, particularly those associated with web shells or communication with known C2 servers (T1071.001) [84].

Furthermore, during the investigation, it was confirmed that hidden files and directories were present, especially in critical system locations (T1564.001) [85]. There are some instances of loadings of non-standard DLLs in IIS servers, particularly those associated with known exploits (e.g., IISCrack.dll) (T1574.002) [86]. Moreover, anomalous financial transactions are observed, deviating slightly from normal patterns, indicating possible fraudulent activity (T1657) [87]. Additionally, there is evidence of data manipulation within financial systems, possibly intended to cover tracks or siphon funds (T1565) [88].

The investigation team identified several patterns using the MITRE ATT&CK [45] framework, but attributing the found traces and attack patterns has become challenging. To address this, the cybersecurity team employed the CAPE tool. By using CAPE tool, analyst can input specific attack patterns mentioned above identified from the breach into the search bar. The tool then generates a list of potential cyber threat actors associated with these patterns, such as FIN10, FIN13 based on CTI documents (discussed in Section 3.4). This capability enables the investigation team to quickly attribute the attack to specific threat actors, providing crucial insights for incident response and mitigation strategies.

### 5.1.2 Scenario: Proactive Threat Hunting

In this section, we discuss a scenario where a cybersecurity team at a healthcare organization is conducting proactive threat hunting activities to identify potential vulnerabilities and threats targeting their data networks. Healthcare facilities are critical infrastructures and contains a huge amount of personal data, where maintaining the privacy of data is very important. Healthcare organizations face numerous potential vulnerabilities and threats due to the sensitive nature of the data they handle, the complexity of their networks, and the increasing sophistication of cyber attackers. During the threat hunting activities, the cybersecurity team found that the use of outdated operating systems and software that are no longer supported with security updates and are still in use in the organization. Furthermore, software on which organization did not apply patches and updates, including electronic health record (EHR) systems, can leave critical vulnerabilities open to exploitation (T1562.010) [89]. In addition, the cybersecurity team observed that there is a lack of proper network segmentation allows attackers to move laterally within the network once they gain access, increasing the risk of widespread damage (T1599) [90].

Several instances are observed in which the organization used default passwords, and failure to implement strong password policies make it easier for attackers to gain unauthorized access (T1589.001) [91]. Additionally, overly permissive user accounts are observed and lack of role-based access controls are noted, which can result in unnecessary exposure of sensitive data [91]. The cybersecurity team also observed that many Internet of Things (IoT) and medical devices lack robust security features, making them easy targets for attackers. They found the use of unsecured communication protocols for device management and data transmission (T1557.003) [92]. The cybersecurity team also identified the third-party vendor risks, where inadequate security measures by third-party vendors can introduce vulnerabilities into the healthcare organization's network (T1195) [93]. The cyber-attacks on supply chain can affect the integrity of the healthcare organization's system. Furthermore, the cybersecurity team noted susceptibility to phishing and social engineering attacks, leading to compromised credentials and unauthorized access. Finally, there is a lack of ongoing training for staff on recognizing and responding to security threats.

The cybersecurity team uses the CAPE tool to explore historical CTI documents and uncover evolving attack patterns and emerging threats. Using the CAPE tool, the cybersecurity team is able to identify the cyber threat actors that are actively launching attacks on the healthcare organizations. Through interactive visualization, they can track the evolution of specific cyber threat actors and their tactics over time. This proactive approach helps the team anticipate and defend against future attacks.

### 5.1.3 Scenario: Integration with Security Information and Event Management (SIEM) Systems

In this section, the effectiveness of our proposed approach is described through a scenario where a large enterprise wants to integrates CAPE with their existing SIEM platform to enhance its threat detection and analysis capabilities. The security information and event management (SIEM) systems often integrate with external threat intelligence feeds that provide information on known threats, indicators of compromise (IOCs), and threat actor tactics, techniques, and procedures (TTPs). This can help in associating observed activities with known threat actors. While SIEM systems are powerful tools for detecting and responding to security incidents, they have limitations when it comes to full threat attribution. They usually lack more detailed and contextual information about threat actors, their motivations, and their past activities, which might not always be available in real-time threat feeds integrated with SIEMs [94], [95]. One important factor is difficult to observe in the conventional SIEM system is the reconstruction of the sequence of events to understand how the attack was carried out. Additionally, insights from human intelligence sources, which can provide information about the intentions and affiliations of threat actors are often lacking [94], [95].

The CAPE tool assists in the aggregation and analysis of unstructured CTI data from various sources. By integrating CAPE with the SIEM platform, the enterprise can conduct in-depth analyses of cyber-attack tactics, techniques, and procedures (TTPs) and attribute them to potential threat actors with high precision. This integration streamlines the identification and response process, reducing the time to detect and mitigate threats.

As CAPE can be utilized to extend the functionality of Security Information and Event Management (SIEM) systems, enabling the cybersecurity analysts to conduct detailed analyses of cyber-attack tactics and attribute them to potential perpetrators with precision. Integration of CAPE with SIEM will enhance the overall cybersecurity framework, offering a comprehensive tool for both pattern analysis and accurate cyber-attack attribution.

### 5.1.4 Scenario: Visualization of Threat Actor Activities

This section describes another scenario where a government agency is monitoring cyber threats and attackers targeting critical infrastructure sectors, such as energy and transportation. Examples of such cyber threat actors targeting transportation are *TA2541, Leviathan, and Tropic Trooper*. The example of groups targeting the energy sector are *Deep Panda, Sandstorm, BITTER, CyberAv3ngers, LAPSUS$, menuPass, Moses Staff, OilRig, Threat Group-3390,* and *Tonto Team*. The agency looks for the attack patterns employed in these sectors by these groups. The CAPE tools provide an interactive platform for visualizing the activities of these cyber threat actors. The cybersecurity analysts can explore and analyze the relationships between different threat actors, their attack patterns, and the impacted sectors. This visualization aids in understanding the broader threat landscape and developing targeted defense strategies.

## 5.2 Implications

This section discusses the implications of our proposed approach presented in this study. Our proposed approach introduces the Cyber Attack Pattern Explorer (CAPE), a practical solution that empowers cybersecurity analysts to identify and explore the tactical cyber-attack patterns.

We introduced a novel approach that combines the non-parametric medoid-seeking clustering technique with the state-of-the-art machine learning models to tackle the challenge of handling large volumes of textual CTI documents. The proposed approach extracts tactical attack patterns, compiling a dataset that serves as the basis for attributing cyber threats. These tactical attack patterns are derived from textual CTI documents, created by human experts based on real cyber-attack incidents, providing valuable insights into cyber threat incidents and threat actors, often surpassing the scope of the currently deployed cybersecurity controls. This approach emphasizes a comprehensive consideration of validity, reliability, generalizability, and overall robustness.

The theoretical implications of this study include the introduction of a hybrid technique that integrates clustering and classification, thereby enriching the theoretical foundations of cyber threat actor attribution. The research highlights the critical role of advanced analytical approaches in shaping cyber threat hunting, emphasizing the imperative for cybersecurity analysts to grasp and adapt to the dynamic landscape of cyber-attack patterns.

The identification and attribution approach presented in this research significantly advances global cybersecurity efforts, offering a robust and accurate solution to cybersecurity analysts. Unlike traditional approaches that often consume considerable time and may lead to inaccuracies, our proposed CAPE tool minimizes the analysis and attribution timeframe. This efficiency empowers cybersecurity analysts, allowing them to promptly identify and report cyber threat incidents, thereby fortifying cybersecurity defenses. This study is aligned with the urgency emphasized by security experts and professionals, reinforcing our commitment to providing timely and reliable support to the cybersecurity community.

The CAPE tool can also serve as an educational resource, empowering students by offering a comprehensive exploration of tactical attack patterns. Through its intuitive search system, students can gain insights into related tactical attack patterns, their temporal evolution, and real-world applications by threat actors. Essentially, the CAPE tool facilitates a narrative-style comprehension of each tactical attack pattern, providing students with a holistic understanding of its nuances and implications.

The study highlights promising avenues for future research, with a particular focus on refining non-parametric techniques tailored for the complexities of textual data, enhancing machine learning models for precise cyber threat actor attribution, and investigating additional dimensions of CTI. A potential enhancement to CAPE could involve integrating attack trace detection mechanisms, enabling cybersecurity analysts to input precise attack artifacts into CAPE's search system. This integration could facilitate autonomous detection of Tactics, Techniques, and Procedures (TTPs), aiding in the identification of specific type of cyber-attacks. Overall, these enhancements would bridge the gap in manual identification of attackers' TTPs and significantly contribute to the automation of cyber threats analysis.

## 6. Conclusion

Navigating the complex landscape of cyber threat attribution remains a difficult challenge in cybersecurity. While machine learning offers automation potential, the scarcity of information on threat actors, and the sophistication of attack methods have limited the contributions to this field. Advanced Persistent Threats (APTs), in particular, stand out as intricate challenges. The increasing number of CTI reports poses an additional hurdle for cybersecurity analysts, given the voluminous yet valuable information they contain. The CAPE emerges as a solution, employing interactive visualization and mining techniques to facilitate efficient information retrieval. A non-parametric mining method is introduced to discern attack patterns from CTI, characterized by semantic similarities with themes commonly employed by cybersecurity analysts. Machine learning algorithms, including Random Forest (RF), Decision Tree (DT), K-Nearest Neighbors (KNN), Deep Learning Neural Network (DLNN), and Naïve Bayes (NB), are trained to attribute cyber threats to their respective actors. Evaluation of this approach encompasses key parameters such as accuracy, precision, recall, F1-score, and false positive rate. The non-parametric approach impressive results: 95.35% accuracy, 95.56% recall, 96.20% precision, 0.93 F1-score, and a 2% false positive rate. The CAPE demonstrates an average precision rate of 88.25% for document retrieval and 88.8% for the list of cyber threat actors. While expert reviews validate the system's applicability, yet acknowledged limitations prompt a call for future enhancements. The framework's dependency on threat data necessitates ongoing refinement, focusing on expanding and improving the quality of the unstructured CTI document corpus. As work progresses, the aim is not merely to expand the corpus size but also to elevate its overall quality.

## Funding Declaration

This research did not receive any specific grant from funding agencies in the public, commercial, or not-for-profit sectors.

## Data Availability

The datasets and source code generated and/or analyzed during the current study are available from the corresponding author upon reasonable request.

## Competing Interests

The authors declare that they have no competing interests.

## Author Contributions (CRediT Statement)

**Conceptualization:** Rimsha Kanwal, Umara Noor

**Methodology:** Rimsha Kanwal, Umara Noor, Zahid Rashid

**Investigation (Experiments, Implementation):** Rimsha Kanwal, Umara Noor

**Literature Review:** Rimsha Kanwal, Umara Noor, Zahid Rashid

**Supervision:** Umara Noor, Zafar Iqbal, Zahid Rashid

**Project Administration:** Umara Noor

**Review & Editing:** Umara Noor, Zahid Rashid

### References

[1] H. Haddadpajouh, A. Azmoodeh, A. Dehghantanha, and R. M. Parizi, "MVFCC: A multi-view fuzzy consensus clustering model for malware threat attribution," *IEEE Access*, vol. 8, pp. 139188–139198, 2020.

[2] Z. Rashid, U. Noor, and J. Altmann, "Economic model for evaluating the value creation through information sharing within the cybersecurity information sharing ecosystem," *Futur. Gener. Comput. Syst.*, vol. 124, pp. 436–466, 2021.

[3] Z. Rashid, U. Noor, and J. Altmann, "Network Externalities in Cybersecurity Information Sharing Ecosystems," Springer, Cham, 2019, pp. 116–125. doi: 10.1007/978-3-030-13342-9_10.

[4] U. Noor, Z. Anwar, J. Altmann, and Z. Rashid, "Customer-oriented ranking of cyber threat intelligence service providers," *Electron. Commer. Res. Appl.*, vol. 41, p. 100976, 2020.

[5] U. Noor, Z. Anwar, T. Amjad, and K.-K. R. Choo, "A machine learning-based FinTech cyber threat attribution framework using high-level indicators of compromise," *Futur. Gener. Comput. Syst.*, vol. 96, pp. 227–242, 2019.

[6] S. Wu, Y. Zhao, F. Parvinzamir, N. T. Ersotelos, H. Wei, and F. Dong, "Literature explorer: effective retrieval of scientific documents through nonparametric thematic topic detection," *Vis. Comput.*, vol. 36, pp. 1337–1354, 2020.

[7] M. Steyvers and T. Griffiths, "Probabilistic topic models," *Handb. latent Semant. Anal.*, vol. 427, no. 7, pp. 424–440, 2007.

[8] S. Deerwester, S. T. Dumais, G. W. Furnas, T. K. Landauer, and R. Harshman, "Indexing by latent semantic analysis," *J. Am. Soc. Inf. Sci.*, vol. 41, no. 6, pp. 391–407, 1990.

[9] T. Hofmann, "Probabilistic latent semantic indexing," in *Proceedings of the 22nd annual international ACM SIGIR conference on Research and development in information retrieval*, 1999, pp. 50–57.

[10] D. M. Blei, A. Y. Ng, and M. I. Jordan, "Latent dirichlet allocation," *J. Mach. Learn. Res.*, vol. 3, no. Jan, pp. 993–1022, 2003.

[11] T. Hofmann, "Probabilistic latent semantic analysis," *arXiv Prepr. arXiv1301.6705*, 2013.

[12] J. Zhang, Y. Song, G. Chen, and C. Zhang, "On-line evolutionary exponential family mixture," in *Twenty-First International Joint Conference on Artificial Intelligence*, 2009.

[13] J. Zhang, Y. Song, C. Zhang, and S. Liu, "Evolutionary hierarchical dirichlet processes for multiple correlated time-varying corpora," in *Proceedings of the 16th ACM SIGKDD international conference on Knowledge discovery and data mining*, 2010, pp. 1079–1088.

[14] D. D. Lee and H. S. Seung, "Learning the parts of objects by non-negative matrix factorization," *Nature*, vol. 401, no. 6755, pp. 788–791, 1999.

[15] J. Kim and H. Park, "Sparse nonnegative matrix factorization for clustering," Georgia Institute of Technology, 2008.

[16] D. Chakrabarti, R. Kumar, and A. Tomkins, "Evolutionary clustering," in *Proceedings of the 12th ACM SIGKDD international conference on Knowledge discovery and data mining*, 2006, pp. 554–560.

[17] P. Ghasiya and K. Okamura, "Investigating Cybersecurity News Articles by Applying Topic Modeling Method," in *2021 International Conference on Information Networking (ICOIN)*, IEEE, 2021, pp. 432–438.

[18] T. Mikolov, I. Sutskever, K. Chen, G. S. Corrado, and J. Dean, "Distributed representations of words and phrases and their compositionality," *Adv. Neural Inf. Process. Syst.*, vol. 26, 2013.

[19] T. Mikolov, K. Chen, G. Corrado, and J. Dean, "Efficient estimation of word representations in vector space," *arXiv Prepr. arXiv1301.3781*, 2013.

[20] J. Pennington, R. Socher, and C. D. Manning, "Glove: Global vectors for word representation," in *Proceedings of the 2014 conference on empirical methods in natural language processing (EMNLP)*, 2014, pp. 1532–1543.

[21] V. Behzadan, C. Aguirre, A. Bose, and W. Hsu, "Corpus and deep learning classifier for collection of cyber threat indicators in twitter stream," in *2018 IEEE International Conference on Big Data (Big Data)*, IEEE, 2018, pp. 5002–5007.

[22] R. Marinho and R. Holanda, "Automated Emerging Cyber Threat Identification and Profiling Based on Natural Language Processing," *IEEE Access*, 2023.

[23] E. Irshad and A. B. Siddiqui, "Cyber threat attribution using unstructured reports in cyber threat intelligence," *Egypt. Informatics J.*, vol. 24, no. 1, pp. 43–59, 2023.

[24] F. Wei *et al.*, "Tiara: a visual exploratory text analytic system," in *Proceedings of the 16th ACM SIGKDD international conference on Knowledge discovery and data mining*, 2010, pp. 153–162.

[25] N. Cao, J. Sun, Y.-R. Lin, D. Gotz, S. Liu, and H. Qu, "Facetatlas: Multifaceted visualization for rich text corpora," *IEEE Trans. Vis. Comput. Graph.*, vol. 16, no. 6, pp. 1172–1181, 2010.

[26] X. Wang, S. Liu, J. Liu, J. Chen, J. Zhu, and B. Guo, "TopicPanorama: A full picture of relevant topics," *IEEE Trans. Vis. Comput. Graph.*, vol. 22, no. 12, pp. 2508–2521, 2016.

[27] W. Dou, X. Wang, R. Chang, and W. Ribarsky, "Paralleltopics: A probabilistic approach to exploring document collections," in *2011 IEEE conference on visual analytics science and technology (VAST)*, IEEE, 2011, pp. 231–240.

[28] W. Cui *et al.*, "Textflow: Towards better understanding of evolving topics in text," *IEEE Trans. Vis. Comput. Graph.*, vol. 17, no. 12, pp. 2412–2421, 2011.

[29] G. Sun, Y. Wu, S. Liu, T.-Q. Peng, J. J. H. Zhu, and R. Liang, “EvoRiver: Visual analysis of topic coopetition on social media,” *IEEE Trans. Vis. Comput. Graph.*, vol. 20, no. 12, pp. 1753–1762, 2014.

[30] O. Thonnard, W. Mees, and M. Dacier, “On a multicriteria clustering approach for attack attribution,” *ACM SIGKDD Explor. Newsl.*, vol. 12, no. 1, pp. 11–20, 2010.

[31] J. Hunker, C. Gates, and M. Bishop, “Attribution requirements for next generation internets,” in *2011 IEEE International Conference on Technologies for Homeland Security (HST)*, IEEE, 2011, pp. 345–350.

[32] K. Geers, D. Kindlund, N. Moran, and R. Rachwald, “World War C: Understanding nation-state motives behind today’s advanced cyber attacks,” *FireEye, Milpitas, CA, USA, Tech. Rep., Sep*, 2014.

[33] E. F. Mejia, “Act and actor attribution in cyberspace: a proposed analytic framework,” *Strateg. Stud. Q.*, vol. 8, no. 1, pp. 114–132, 2014.

[34] P. Shakarian, G. I. Simari, G. Moores, and S. Parsons, “Cyber attribution: An argumentation-based approach,” *Cyber Warf. Build. Sci. Found.*, pp. 151–171, 2015.

[35] F. Stranne, U. Bilstrup, and L. Ewertsson, “Behind the Mask–Attribution of antagonists in cyberspace and its implications on international conflicts and security issues,” in *International Studies Association (ISA)’s 56th Annual, Convention–Global IR and Regional Worlds. A New Agenda for International Studies, New Orleans, Louisiana, United States, February 18-21, 2015*, 2015.

[36] B. Edwards, A. Furnas, S. Forrest, and R. Axelrod, “Strategic aspects of cyberattack, attribution, and blame,” *Proc. Natl. Acad. Sci.*, vol. 114, no. 11, pp. 2825–2830, 2017.

[37] H. Mwiki, T. Dargahi, A. Dehghantanha, and K.-K. R. Choo, “Analysis and triage of advanced hacking groups targeting western countries critical national infrastructure: Apt28, red october, and regin,” *Crit. Infrastruct. Secur. Resil. Theor. Methods, Tools Technol.*, pp. 221–244, 2019.

[38] MITRE ATT&CK, “MITRE ATT&CK®.” https://attack.mitre.org/ (accessed Jan. 13, 2024).

[39] A. Warikoo, “The triangle model for cyber threat attribution,” *J. Cyber Secur. Technol.*, vol. 5, no. 3–4, pp. 191–208, 2021.

[40] S. Naveen, R. Puzis, and K. Angappan, “Deep learning for threat actor attribution from threat reports,” in *2020 4th International Conference on Computer, Communication and Signal Processing (ICCCSP)*, IEEE, 2020, pp. 1–6.

[41] L. Perry, B. Shapira, and R. Puzis, “No-doubt: Attack attribution based on threat intelligence reports,” in *2019 IEEE International Conference on Intelligence and Security Informatics (ISI)*, IEEE, 2019, pp. 80–85.

[42] A. Sentuna, A. Alsadoon, P. W. C. Prasad, M. Saadeh, and O. H. Alsadoon, “A novel Enhanced Naïve Bayes Posterior Probability (ENBPP) using machine learning: Cyber threat analysis,” *Neural Process. Lett.*, vol. 53, pp. 177–209, 2021.

[43] N. Dionísio, F. Alves, P. M. Ferreira, and A. Bessani, “Cyberthreat detection from twitter using deep neural networks,” in *2019 international joint conference on neural networks (IJCNN)*, IEEE, 2019, pp. 1–8.

[44] B. D. Le, G. Wang, M. Nasim, and A. Babar, “Gathering cyber threat intelligence from Twitter using novelty classification,” *arXiv Prepr. arXiv1907.01755*, 2019.

[45] MITRE ATT&CK, “MITRE ATT&CK®”.

[46] Google, “Programmable Search Engine.” https://cse.google.com/cse?cx=003248445720253387346:turlh5vi4xc (accessed Jul. 07, 2024).

[47] K. Bandla, “APT Notes.” https://github.com/kbandla/APTnotes (accessed Jul. 27, 2022).

[48] BrightTALK, “Discover and learn with the world’s brightest professionals.”

[49] Feedspot Blog, “Top 20 Cyber Security Forums in 2024.” https://forums.feedspot.com/cyber_security_forums/ (accessed Jan. 27, 2024).

[50] MalwareTips, “Malware Tips Forum.” https://malwaretips.com/ (accessed Jan. 27, 2024).

[51] Wilder Security Forum, “Wilder Security Forum.” https://www.wilderssecurity.com/ (accessed Jan. 27, 2024).

[52] IT Security Guru, “IT Security Guru.” https://www.itsecurityguru.org/ (accessed Jan. 27, 2024).

[53] SC Media, “SC Media.” https://www.scmagazine.com/security-weekly-blog (accessed Jan. 27, 2024).

[54] The Hacker News, “The Hacker News - Most Trusted Cyber Security and Computer Security Analysis.” https://thehackernews.com/ (accessed Jan. 27, 2024).

[55] I. Magazine, “Infosecurity Magazine - Strategy, Insight, Technology.” https://www.infosecurity-magazine.com/ (accessed Jan. 27, 2024).

[56] CSO, “CSO - Latest from today.” https://www.csoonline.com/ (accessed Jan. 27, 2024).

[57] Helpnetsecurity, “Help Net Security.” https://www.helpnetsecurity.com/ (accessed Jan. 27, 2024).

[58] Techtarget, “SearchSecurity.” https://www.techtarget.com/searchsecurity/ (accessed Jan. 27, 2024).

[59] ThreatMiner.org, “ThreatMiner.org - Data Mining for Threat Intelligence,.” https://www.threatminer.org/ (accessed Jan. 20, 2024).

[60] CISA, “CISA, CERT.” https://www.cisa.gov/uscert/ (accessed Jan. 27, 2024).

[61] Palo Alto Networks, “Global Cybersecurity Leader - Palo Alto Networks.” https://www.paloaltonetworks.com/ (accessed Jan. 27, 2024).

[62] CISCO, “CISCO.” https://www.cisco.com/site/us/en/products/security/index.html (accessed Jan. 27, 2024).

[63] IBM, “Key cybersecurity technologies.” https://www.ibm.com/topics/cybersecurity (accessed Jan. 27, 2024).

[64] CrowdStrike, “Crowdstrike: Stop breaches. drive business.” https://www.crowdstrike.com (accessed Jan. 13, 2024).

[65] Secureworks, “Secureworks: Cybersecurity leader, proven threat defense.”

[66] FireEye, “Cyber security experts solution providers.”

[67] I. Mike Bostock and Observable, “D3js - The JavaScript library for bespoke data visualization.” https://d3js.org/ (accessed Jul. 13, 2024).

[68] django, “Django - The web framework for perfectionists with deadlines.” https://www.djangoproject.com/ (accessed Jul. 13, 2024).

[69] MITRE ATT&CK, “Exploit Public-Facing Application.” https://attack.mitre.org/techniques/T1190/ (accessed Jul. 05, 2024).

[70] MITRE ATT&CK, “External Remote Services.” https://attack.mitre.org/techniques/T1133/ (accessed Jul. 05, 2024).

[71] MITRE ATT&CK, “OS Credential Dumping: LSASS Memory.” https://attack.mitre.org/techniques/T1003/001/ (accessed Jul. 05, 2024).

[72] MITRE ATT&CK, “OS Credential Dumping: Security Account Manager.” https://attack.mitre.org/techniques/T1003/002/ (accessed Jul. 05, 2024).

[73] MITRE ATT&CK, “Network Service Discovery.” https://attack.mitre.org/techniques/T1046/ (accessed Jul. 05, 2024).

[74] MITRE ATT&CK, “Boot or Logon Autostart Execution: Registry Run Keys / Startup Folder.” https://attack.mitre.org/techniques/T1547/001/ (accessed Jul. 05, 2024).

[75] MITRE ATT&CK, “Account Manipulation.” https://attack.mitre.org/techniques/T1098/ (accessed Jul. 05, 2024).

[76] MITRE ATT&CK, “Command and Scripting Interpreter: PowerShell.” https://attack.mitre.org/techniques/T1059/001/ (accessed Jul. 05, 2024).

[77] MITRE ATT&CK, “Scheduled Task/Job: Scheduled Task.” https://attack.mitre.org/techniques/T1053/005/ (accessed Jul. 05, 2024).

[78] MITRE ATT&CK, “Remote Services: Remote Desktop Protocol.” https://attack.mitre.org/techniques/T1021/001/ (accessed Jul. 05, 2024).

[79] MITRE ATT&CK, “Remote Services: SMB/Windows Admin Shares.” https://attack.mitre.org/techniques/T1021/002/ (accessed Jul. 05, 2024).

[80] MITRE ATT&CK, “Remote Services: SSH.” https://attack.mitre.org/techniques/T1021/004/ (accessed Jul. 05, 2024).

[81] MITRE ATT&CK, “Remote Services: Windows Remote Management.” https://attack.mitre.org/techniques/T1021/006/ (accessed Jul. 05, 2024).

[82] MITRE ATT&CK, “Data Staged: Local Data Staging.” https://attack.mitre.org/techniques/T1074/001/ (accessed Jul. 05, 2024).

[83] MITRE ATT&CK, “Archive Collected Data: Archive via Utility.” https://attack.mitre.org/techniques/T1560/001/ (accessed Jul. 05, 2024).

[84] MITRE ATT&CK, “Application Layer Protocol: Web Protocols.” https://attack.mitre.org/techniques/T1071/001/ (accessed Jul. 05, 2024).

[85] MITRE ATT&CK, “Hide Artifacts: Hidden Files and Directories.” https://attack.mitre.org/techniques/T1564/001/ (accessed Jul. 05, 2024).

[86] MITRE ATT&CK, “Hijack Execution Flow: DLL Side-Loading.” https://attack.mitre.org/techniques/T1574/002/ (accessed Jul. 05, 2024).

[87] MITRE ATT&CK, “Financial Theft.” https://attack.mitre.org/techniques/T1657/ (accessed Jul. 24, 2024).

[88] MITRE ATT&CK, “Data Manipulation.” https://attack.mitre.org/techniques/T1565/ (accessed Jul. 05, 2024).

[89] MITRE ATT&CK, “Impair Defenses: Downgrade Attack.” https://attack.mitre.org/techniques/T1562/010/

[90] MITRE ATT&CK, “Network Boundary Bridging.” https://attack.mitre.org/techniques/T1599/ (accessed Jul. 05, 2024).

[91] MITRE ATT&CK, “Gather Victim Identity Information: Credentials.” https://attack.mitre.org/techniques/T1589/001/ (accessed Jul. 05, 2024).

[92] MITRE ATT&CK, “Adversary-in-the-Middle: DHCP Spoofing.” https://attack.mitre.org/techniques/T1557/003/ (accessed Jul. 05, 2024).

[93] MITRE ATT&CK, “Supply Chain Compromise.” https://attack.mitre.org/techniques/T1195/ (accessed Jul. 06, 2024).

[94] S. Saeed, S. A. Suayyid, M. S. Al-Ghamdi, H. Al-Muhaisen, and A. M. Almuhaideb, “A systematic literature review on cyber threat intelligence for organizational cybersecurity resilience,” *Sensors*, vol. 23, no. 16, p. 7273, 2023.

[95] SOCRadar, “Harnessing SIEM Solutions With Threat Intelligence.” https://socradar.io/harnessing-siem-solutions-with-threat-intelligence/ (accessed Jul. 06, 2024).